%% file: main.tex
\documentclass[sigconf]{acmart}
\PassOptionsToPackage{prologue,dvipsnames}{xcolor} 
\usepackage{subfiles}
\usepackage{makecell}
\usepackage{enumitem}
\usepackage{multirow}
\usepackage{longtable}
\usepackage{graphicx}
\usepackage{caption}
\usepackage{subcaption}
\usepackage{tabularx}
\usepackage{soul}
\usepackage{float}
\definecolor{darkgreen}{rgb}{0.0, 0.5, 0.0}
\definecolor{navyblue}{rgb}{0.0, 0.0, 0.5}

\newif\ifredact
\newif\ifcomment

\redactfalse  
\commenttrue  

\newcommand{\dquote}[1]{\textit{``#1''}}

\ifcomment
  \newcommand{\missing}[1]{\textcolor{red}{~#1}}
  \newcommand{\ken}[1]{~\sethlcolor{yellow}\hl{[Kenny: #1]}}
  \newcommand{\wei}[1]{~\sethlcolor{pink}\hl{[Weiyan: #1]}}
  \newcommand{\discuss}[1]{~\sethlcolor{green}\hl{[Used in discussion: #1]}}
\else
  \newcommand{\missing}[1]{}
  \newcommand{\ken}[1]{}
  \newcommand{\wei}[1]{}
  \newcommand{\discuss}[1]{}
\fi

\AtBeginDocument{%
  }

\setcopyright{acmlicensed}
\copyrightyear{2018}
\acmYear{2018}
\acmDOI{XXXXXXX.XXXXXXX}

\acmConference[Conference acronym 'XX]{Make sure to enter the correct
  conference title from your rights confirmation emai}{June 03--05,
  2018}{Woodstock, NY}
\acmISBN{978-1-4503-XXXX-X/18/06}

\begin{document}
\renewcommand\footnotetextcopyrightpermission[1]{} 
\pagestyle{plain} 
\settopmatter{printacmref=false} 

\title{Human–AI Alignment of Multimodal Large Language Models with Speech-Language Pathologists in Parent–Child Interactions}

\author{Weiyan Shi}
\email{weiyanshi6@gmail.com}
\orcid{0009-0001-6035-9678}
\affiliation{
  \institution{Singapore University of Technology and Design}
  \country{Singapore}
}

\author{Kenny Tsu Wei Choo}
\email{kennytwchoo@gmail.com}
\orcid{0000-0003-3845-9143}
\affiliation{
  \institution{Singapore University of Technology and Design}
  \country{Singapore}
}
\authornote{Corresponding author. This preprint corresponds to an earlier version of the work released in May 2025. 
The version accepted at CHI 2026 is available as a separate preprint at https://arxiv.org/abs/2511.04366.}

\renewcommand{\shortauthors}{Shi et al.}
\input{sections/0_abstract}


\begin{teaserfigure}
  \centering
  \includegraphics[width=0.4\textwidth]{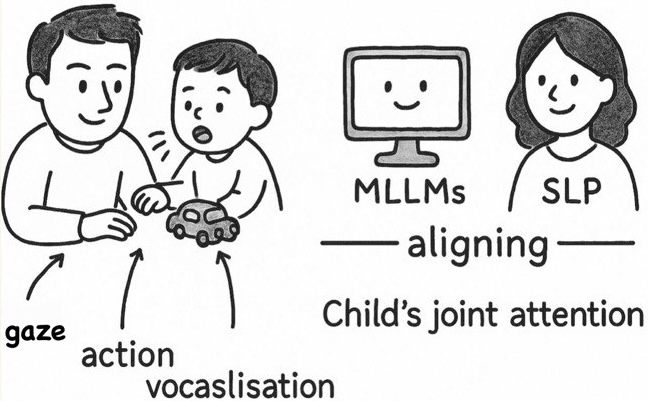}
  \caption{Conceptual overview: Our work explores how MLLMs can be guided to observe and judge parent–child interactions like SLPs, using structured behavioural cues such as gaze, action, and vocalisation.}
  \Description{Diagram showing how MLLMs observe and judge child–parent interaction using gaze, action, and vocalisation.}
  \label{fig:banner}
\end{teaserfigure}


\maketitle


\input{sections/1_introduction}
\input{sections/2_relatedwork}
\input{sections/3_pci_video_dataset}
\input{sections/4_slp_interview}
\input{sections/5_mllm_system}
\input{sections/6_design_guide}
\input{sections/7_discussion}
\input{sections/8_conclusion}


\bibliographystyle{ACM-Reference-Format}
\bibliography{main}

\appendix

\end{document}
\end{document}

%% file: sections/0_abstract.tex
\begin{abstract}
    Joint attention is a critical marker of early social-communicative development, yet remains difficult for caregivers to assess without expert guidance. In this work, we explore how multimodal large language models (MLLMs) can be aligned with the reasoning processes of speech-language pathologists (SLPs) to support the interpretation of everyday parent–child interactions. We conducted in-depth interviews and video annotation studies with three experienced SLPs to uncover how they evaluate joint attention based on three core behavioural cues: gaze, action, and vocalisation. Using these insights, we developed a two-stage MLLM-based system that first extracts fine-grained behavioural descriptions from video segments and then judge joint attention quality using expert-aligned prompts. Our evaluation across 26 parent-child interaction videos shows that MLLMs can achieve up to 85\% accuracy in perceptual cue extraction and over 75\% average precision in simulating expert judgement. We further propose design guidelines for building MLLM-based behaviour observation-judgement systems that align with SLPs, emphasising the structuring of behavioural cues, the construction of exemplar libraries grounded in expert annotations, and the need to personalise system responses based on developmental stage and neurotypical or atypical presentation. This work provides structured behavioural cues derived from SLP expertise, demonstrates the feasibility of aligning SLPs observation and judgement using MLLMs, and offers practical design guidelines for building aligned systems to support parent–child interaction analysis.
\end{abstract}

%% file: sections/1_introduction.tex
\section{Introduction}
Despite recent advances in artificial intelligence (AI) within family and education~\cite{yuan2024designing, sun2024exploring, fiani2024exploring, nikkhah2024family, currin2024opportunities, su2024hidden}, existing systems fall short in comprehending early parent–child interactions in a manner consistent with expert clinical observing and judging. 
This shortfall is critical, considering that approximately 1 in 6 children in the United States experience at least one developmental delay~\cite{robinson2017cdc}, with language delays being among the most prevalent~\cite{sunderajan_speech_2019}.

Foundational social-communicative behaviours such as joint attention, joint intention, and social referencing often emerge during play.
However, these behaviours are subtle and challenging for parents to identify without professional guidance.
Speech language pathologists (SLPs) observe these behaviours in a multimodal observations--through gaze, gestures, vocalisations, and contextual information.
Yet, access to SLP services is often hindered by factors such as high costs, time constraints, and limited availability, particularly in rural or underserved areas~\cite{o2005barriers}.
As a result, expert developmental knowledge and care remains largely out of reach in everyday caregiving contexts.


Current AI research in the speech-language therapy domain has largely focused on supporting SLPs or caregivers, rather than simulating expert observing and judging itself.
Lewis et al.~\cite{lewis2025exploring} investigated how SLPs interact with AI-generated materials for children from culturally and linguistically diverse backgrounds. Their findings revealed concerns around representativeness, bias, and usability, suggesting that current tools lack contextual awareness. However, their work focuses on evaluating content usability, not on enabling AI to replicate the evaluative criteria of SLPs.
On the caregiver side, Dangol et al.~\cite{dangol2025want} explored how parents can be supported in delivering speech therapy techniques at home. Their study identified common emotional and logistical challenges and proposed AI design concepts to assist in engagement and adherence. Yet their system assumes that expert evaluation still resides with the SLP, and does not provide mechanisms for interpreting children’s behaviours in the way an expert might.
However, there is a lack of research addressing whether AI systems can interpret everyday parent-child interactions in a manner that aligns with expert clinical observing and judging process.
In our research, we posit that AI has the potential to serve as an initial reflective support tool, offering parents insights even before formal intervention is sought--provided it can reliably identify key indicators and \emph{emulate expert-like interpretation}.
This also provides an evidenced manner to approach care with SLPs and/or to aid diagnosis beyond parent self-report and less frequent SLP observations.

In evaluating complex social behaviours such as joint attention, SLPs rely on holistic judgments informed by gaze, gesture, vocalisation, and context. 
While grounded in developmental theory, these evaluations are typically applied through intuitive, experience-driven observing and judging process that is difficult to formalise or quantify. 
This presents a challenge for AI: although multimodal large language models (MLLMs) excel at well-defined tasks like video captioning or temporal grounding, they struggle with socially nuanced, subjective tasks that lack clear ground truth. 
Without alignment to SLP observing and judging process, MLLMs remain limited in their ability to simulate clinical-level assessments in parent–child interactions.

In this study, we take joint attention as a representative case to explore how MLLMs can be aligned with the observing and judging process of SLPs in analysing parent–child interactions. While joint attention is a well-established marker of interaction quality, SLPs typically evaluate it through integrated and intuitive interpretations of behavioural cues—-such as gaze, vocalisation, and action—-which are difficult to formalise. Our goal is to understand how these expert criteria can be translated into structured formats that MLLMs can interpret and execute, enabling alignment between model outputs and clinical observing and judging process. 

Guided by this objective, we address the following research questions (RQs):

\begin{itemize}
    \item \textbf{RQ1: How do SLPs observe and analyse parent–child interactions?} What criteria do they use to identify segments of strong or poor joint attention? How consistent are their judgments?
    \item \textbf{RQ2: How can we represent SLPs’ evaluative criteria in formats that MLLMs can both understand and execute, and what design choices best support alignment between MLLM and expert observation and judgement?}
\end{itemize}

To investigate these questions, we conducted in-depth interviews and annotation sessions with three experienced SLPs, who annotated 25 selected videos of parent–child interaction for strong and poor joint attention. Through this process, we identified three key behavioural dimensions—\textit{gaze}, \textit{vocalisation}, and \textit{action}—that underlie SLP judgments. We then translated these heuristics into structured, MLLM-compatible task formats: first prompting models to extract behavioural segments from raw video, and then prompting them to assess interaction quality using few-shot examples.

\textbf{Our key contributions are:}
\begin{itemize}
    \item We introduce a three-dimensional framework for evaluating joint attention in parent–child interactions, grounded in expert interviews and annotation analysis. This framework decomposes expert judgement into three core behavioural cues—\textbf{gaze}, \textbf{action}, and \textbf{vocalisation}—offering a structured lens through which MLLMs can observe and judge joint attention.

    \item We develop an expert-aligned MLLM-based system that simulates SLP-style joint attention assessment in two stages: (1) observing fine-grained behavioural cues from parent-child interaction videos using expert-informed prompting, achieving up to 85\% accuracy across dimensions; and (2) evaluating interaction quality using only structured behavioural descriptions, reaching over 75\% average precision compared to expert labels.

    \item We curate a segment-level dataset with expert-labelled joint attention behaviours and derive a set of practical design guidelines for building MLLM-based systems that align SLP’s observing and judging process. These guidelines cover prompt engineering, cue structuring, model configuration, and future directions for parent-facing AI systems.
\end{itemize}

%% file: sections/2_relatedwork.tex
\section{RELATED WORK}
\subsection{Language Delay and the Limits of Current Parent Support}

Language delay is one of the most common developmental concerns in early childhood, with prevalence estimates ranging from 2.3\% to 19\% among children aged 2 to 7 years~\cite{mclaughlin2011speech}. Without timely intervention, children with language delay are at increased risk for later difficulties in reading, attention regulation, and social interaction~\cite{sunderajan2019speech}. SLPs play a key role in identifying and supporting children with language delays through structured interventions. However, given limited access to professional services in many settings, recent efforts have increasingly focused on empowering parents to support their children at home~\cite{lieneman2017parent}.
In Singapore, access to publicly funded speech therapy services is often limited by long wait times, and private therapy can be prohibitively expensive for many families~\cite{straitstimes2024earlyintervention}. As a result, empowering caregivers to support early communication at home has become a practical necessity.

Parent training programmes such as \textit{It Takes Two to Talk} and \textit{More Than Words} have shown success in encouraging behaviours that promote early language development~\cite{pepper2004talk, sussman1999words, brock2014statewide, finke2009all, gadberry2011survey, ganz2013impacts}. These interventions typically teach caregivers how to observe, wait, and respond during everyday interactions. Yet while such strategies may improve the quantity and quality of parent–child communication, they rarely help caregivers evaluate the developmental effectiveness of those interactions.

In particular, parents lack support for recognising nuanced social-communicative milestones such as joint attention—an early and powerful predictor of later language and cognitive development. Unlike general verbal encouragement or play-based engagement, joint attention requires moment-to-moment sensitivity to the child’s gaze, gestures, and responses. Evaluating whether joint attention is present—and whether it is strong, partial, or absent—often demands the contextual and multimodal judgement that experienced SLPs bring~\cite{mundy2007individual}. Unfortunately, families outside clinical environments rarely have access to such expert judgment~\cite{roberts2011parent}. This gap motivates our exploration of AI-assisted systems that can simulate SLP judgement and offer parents reflective insights into their interactions—support that goes beyond behaviour modelling toward meaningful, expert-aligned interpretation.

\subsection{Evaluating Joint Attention: Limits of Parent Training and the Role of Expert Judgment}
Joint attention refers to the shared focus between a child and caregiver on the same object or event, typically established through gaze, gestures such as pointing, and verbal cues~\cite{sussman1999words}. It is one of the earliest and most important indicators of healthy social-communicative development, and difficulties in joint attention are often early signs of developmental delay~\cite{mundy2007individual, tomasello1986joint}. Several parent-oriented intervention programmes—such as \textit{DIR Floortime}~\cite{dir_floortime}, \textit{It Takes Two to Talk}~\cite{pepper2004talk}, and \textit{More Than Words}~\cite{sussman1999words}—have been developed to help caregivers guide and encourage joint attention behaviours through emotionally attuned and child-led interaction. However, while these programmes can help parents learn how to promote joint attention, they do not equip them to evaluate its quality. Accurately assessing whether joint attention is strong, weak, or absent often requires the nuanced judgment of experienced SLPs, who integrate cues such as gaze, vocalisation, and action into holistic, context-sensitive assessments. These evaluations are typically based on years of clinical experience and remain difficult to formalise into explicit rules. As a result, families must rely on expert observation—yet access to SLPs is often limited due to high cost, time constraints, and service shortages, particularly in non-clinical or underserved settings~\cite{o2005barriers}. This gap highlights a pressing need for scalable, AI-assisted tools that can support the identification and assessment of joint attention in naturalistic parent–child interactions.

\subsection{Technological Support for Parent-Child Interaction}
Recent years have seen a surge in interactive technologies designed to support early childhood development, with a particular emphasis on enhancing the quality and frequency of parent–child interaction.
Chan et al. developed WAKEY, a system designed to improve parent-child communication during morning routines, enabling parents and preschool children to start their day more smoothly~\cite{chan2017wakey}. However, WAKEY relies entirely on manual input and logging by parents for tasks such as scheduling and tracking phrase usage frequency, lacking the ability to automatically gather contextual data, such as audio or video interaction details.
Hwang et al. proposed \textit{TalkBetter}, a system that analyses turn-taking and provides real-time feedback and tailored recommendations to help parents foster their children's language development~\cite{hwang2014talkbetter}.
Song et al. introduced \textit{TalkLIME}, a mobile system that enhances parent-child interactions by providing real-time feedback on metrics like utterance count, turn-taking, and initiation ratio~\cite{song2016talklime}.
Jeong et al. developed \textit{Captivate!}, a system that uses multimodal sensing, including gaze estimation and speech recognition, to detect joint attention and recommend phrases during parent-child interactions~\cite{kwon2022captivate}.
Choi et al. developed \textit{AACessTalk}, a tablet-based system that supports communication between minimally verbal autistic children and their parents by providing contextual guidance and vocabulary card recommendations. Through a two-week deployment study, they demonstrated how the system improved turn-taking, increased communication frequency, and helped parents reflect on their interaction strategies~\cite{choi2025aacess}.
Dangol et al.~\cite{dangol2025want} explored how parents can be supported in learning and applying speech therapy techniques at home. Their work focuses on helping caregivers understand and adopt expert-recommended strategies, identifying common emotional and logistical barriers and proposing AI design concepts to support engagement and adherence. However, the system remains a support tool centred on strategy delivery, and does not attempt to interpret children’s behaviours or assess interaction quality from an expert perspective.

However, these systems and methods primarily serve as design interventions to scaffold or enhance parent–child interaction, rather than tools for behavioural observation and judgement. None provide parents with expert-aligned evaluations of interaction quality from the perspective of SLPs, who rely on nuanced, context-dependent behavioural cues—such as gaze, turn-taking, and affective engagement—to assess social-communicative development.

\subsection{Technological Support for SLPs}
Generative AI is beginning to influence many aspects of clinical practice, including how SLPs source or adapt therapy materials. For example, Lewis et al.~\cite{lewis2025exploring} examined how SLPs interact with AI-generated visual content when working with children from culturally and linguistically diverse backgrounds. While their study highlighted important concerns regarding representation, bias, and contextual mismatch, it focused primarily on content usability. Their work does not address how AI might support or simulate the evaluative reasoning processes that SLPs use when observing and interpreting children’s behaviours in real time.
With the rapid progress of MLLMs across tasks such as video captioning and temporal grounding~\cite{lu2024gpt}, researchers have begun to explore their potential for analysing human behaviour—-particularly in socially grounded contexts such as parent–child interaction. MLLMs can interpret complex verbal and nonverbal behaviour across video, audio, and text, and generate descriptive summaries of human activity~\cite{gandhi2023multimodal, jain2024vcoder}. Thanks to their in-context learning capabilities, MLLMs can flexibly adapt to new interaction settings with minimal data~\cite{dos2023composite,whitehead2024generative}, making them promising for contexts that require domain-specific reasoning. Crucially, these models shift the output from fixed metrics or visualisation into natural language descriptions—potentially bridging the gap between raw behavioural signals and expert interpretation.
Zheng et al.~\cite{zheng2024soap} enables experts to interact with MLLMs through in-context prompting and collaborative task design, and supports the automatic generation of initial behavioural segments—such as those relevant to joint attention—in domains like healthcare and education. However, SOAP.AI is designed primarily as a tool for expert users, and does not evaluate the reliability of its outputs or align them with SLP judgement. Its goal is to augment expert workflows, not to investigate whether MLLMs can simulate expert evaluative capabilities.

While these approaches centre on supporting SLPs in their clinical workflows, they do not address a critical question: can current technologies simulate the observational and interpretive reasoning that SLPs apply when evaluating parent–child interactions? Bridging this gap requires rethinking how AI systems can align with expert judgement—not just to assist SLPs, but to empower parents with meaningful, expert-level insights into their everyday interactions.

\subsection{Towards SLP-Aligned Observation and Judgement in Parent–Child Interaction}

The first attempt to align MLLMs with the evaluative judgement of SLPs was introduced by Shi et al.~\cite{shi2025towards}, who prompted MLLMs with definitions of joint attention to directly identify strong or poor joint attention segments through a temporal grounding task. Their results showed poor performance, primarily due to the models’ limited sensitivity to fine-grained gaze cues—particularly eye contact, which is central to joint attention dynamics. These findings highlight the need for more structured alignment between MLLM interpretation and the evaluative judgement used by SLPs, rather than relying solely on definition-based prompting.

Our work builds on Shi et al.~\cite{shi2025towards} by moving beyond temporal grounding to examine how MLLMs can be aligned with the evaluative judgement of SLPs. Rather than relying on raw video inputs alone, we introduce a structured, heuristic-based framework that translates expert judgement criteria into promptable representations. This enables MLLMs not only to identify relevant behavioural segments, but also to simulate expert-like assessments of interaction quality—opening new possibilities for AI-assisted support in naturalistic, non-clinical settings. We demonstrate this approach in the context of parent–child interaction, using the concept of \textit{joint attention}—a foundational marker of social-communicative development—as a test case for expert-aligned interpretation.

%% file: sections/3_pci_video_dataset.tex
\section{Video Dataset of Parent-Child Interaction}

To support both the analysis of how SLPs observe and judge joint attention (RQ1) and the development of MLLM-based methods aligned with these expert approaches to observation and judgement (RQ2), we curated a central dataset of parent–child interaction videos.
For RQ1, the dataset was intentionally designed to span a range of child age groups and interaction contexts, allowing us to probe how SLPs observe and judge strong versus poor joint attention behaviours across developmental stages. By including diverse activities—such as structured tasks, free play, and instructional moments—we also gathered expert input on which types of interaction best support joint attention observation and judgement.
For RQ2, we used the same dataset to develop and validate an MLLM-based system aligned with SLP observation and judgement. Specifically, we prompted the model to (1) observe key behavioural cues in each segment and (2) judge their joint attention quality using structured guidance derived from SLP annotations. This dataset thus served as a shared benchmark for analysing expert observation and judgement, and for evaluating how well MLLM outputs align with SLP approaches to observing and judging parent–child interactions (see Section~\ref{sec:slp_interview} for RQ1 and Section~\ref{sec:mllm_system} for RQ2).

\subsection{Video Selection Method}
We collected videos from YouTube using targeted search terms aimed at capturing naturalistic parent–child interactions. Keywords included phrases such as \textit{“parent–child play,” “playing with toddler,” “interactive play with baby,” “floor time with child,”} and \textit{“talk to your baby”}. Our goal was to find realistic, continuous recordings that retained the natural beginning, middle, and end of an interaction—crucial for observing how joint attention behaviours emerge and evolve. We excluded short or selectively edited clips, which often lacked the contextual richness necessary for such analysis.

To maintain a clear analytical focus, each video was required to depict a single adult and a single child engaged in dyadic interaction. We prioritised unscripted, organically unfolding exchanges over demonstration or training footage, ensuring ecological validity for our proof-of-concept study.

We selected videos that offered sufficiently clear views of both the child and adult to support basic analysis of gaze, gesture, and vocalisation. While we did not require professional production quality, we favoured videos with stable framing and minimal camera movement to reduce ambiguity in interpreting visual cues. Adequate audio quality was also important to allow for general recognition of vocal exchanges, though minor background noise was acceptable given the naturalistic setting.

Additionally, the dataset was curated to include children across a broad age range to capture a variety of developmental stages. We prioritised videos that showcased diverse and meaningful interactions, such as language learning activities, skill-building tasks, or natural daily exchanges. All textual elements, including subtitles and transitions, were removed to preserve the raw multimodal content.

\subsection{Dataset Composition and Analysis}
We curated a final dataset of 26 parent–child interaction videos that span key developmental stages from infancy to early school age (0–8 years), with most videos focusing on the critical preschool years. The age information was obtained from publicly available video descriptions. Specifically, the dataset includes 4 videos featuring children aged 0–2, 5 videos for ages 2–4, 16 videos for ages 4–6, and 1 video featuring children aged 6–8. These videos were created and shared by parent–child interaction experts who have published numerous resources aimed at modelling effective interaction techniques and communication strategies for caregivers.

The duration of the videos ranges from brief clips of around 0.5 to 12 minutes. Most videos are relatively short: 13 videos are less than 1 minute, and 11 fall between 1 and 5 minutes. Only two videos extend beyond 5 minutes, one categorised as 5–10 minutes and the other over 10 minutes. This distribution highlights the concise nature of the interactions and tasks typically observed in parent-child settings.

The video content in the dataset can be categorised into three main categories. The first category is behaviour guidance and skill modelling\footnote{\url{https://www.youtube.com/watch?v=YUkujhg6j6w}} (n=10), where parents demonstrate specific techniques (e.g., \textit{PRIDE skills}~\cite{masse2018taking}, \textit{\dquote{Big Ignore} techniques})~\cite{woodfield2021time} to guide children's behaviour and improve interaction quality. The second category focuses on language and cognitive development\footnote{\url{https://www.youtube.com/watch?v=rVqJacvywAQ}} (n=10), showcasing how parent-child interactions foster language learning and cognitive skills through tasks or experiments, such as discussing specific topics or engaging in Piaget's cognitive experiments. The third category involves daily life skills and interaction\footnote{\url{https://www.youtube.com/watch?v=N3wAPLXd7I0}} (n=6), presenting natural exchanges in structured settings, such as reviewing reward charts or ending special playtime, emphasising practical life skills and building strong parent-child relationships.

\begin{figure}[ht]
    \centering
    \begin{subfigure}[t]{0.31\textwidth}
        \includegraphics[width=\linewidth]{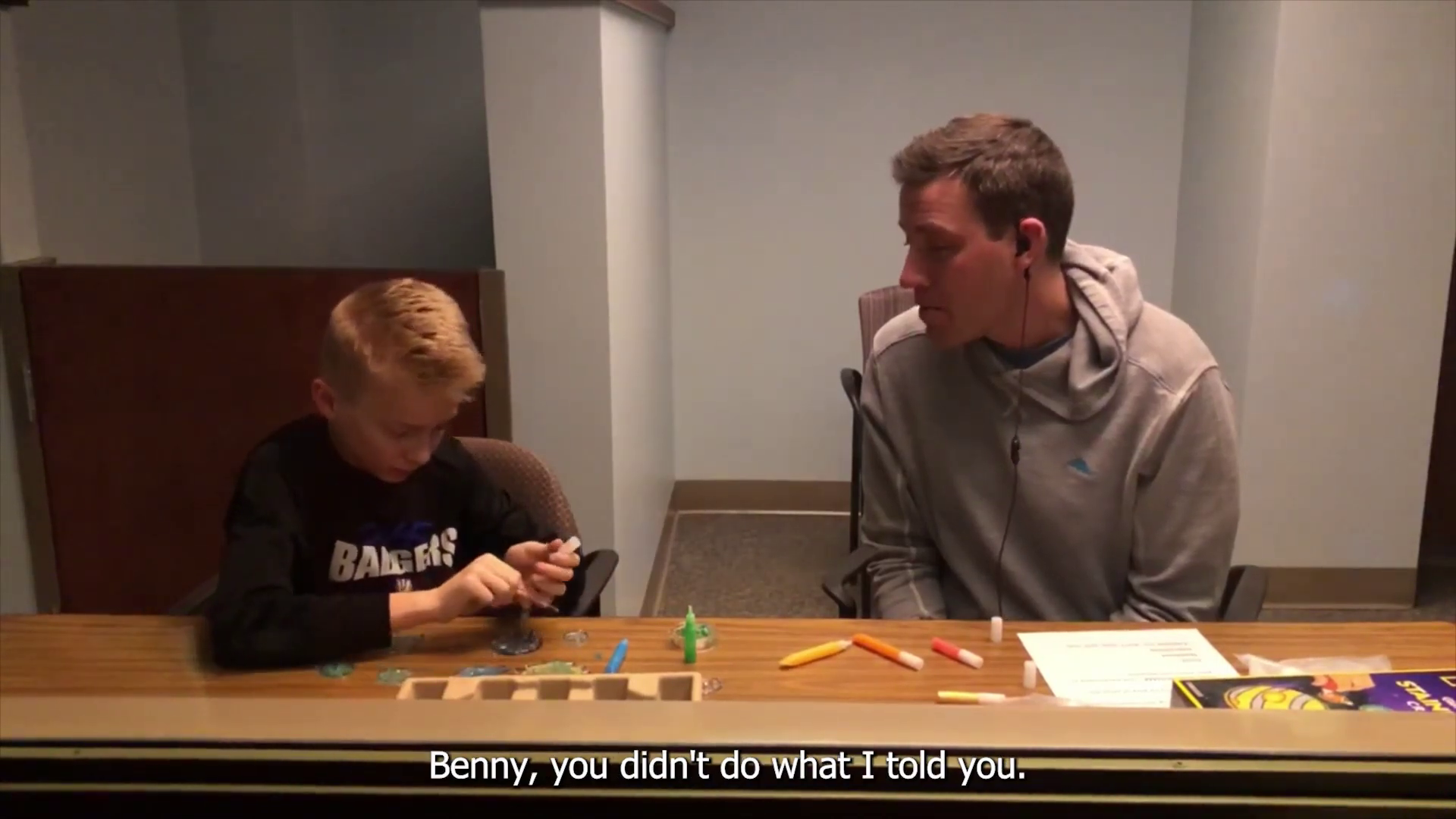}
        \caption{\textbf{Behavioural Guidance and Skill Modelling}: 
        Parent uses the \textit{"Big Ignore"}~\cite{woodfield2021time} technique while the child continues painting.}
    \end{subfigure}
    \hfill
    \begin{subfigure}[t]{0.31\textwidth}
        \includegraphics[width=\linewidth]{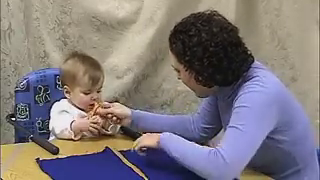}
        \caption{\textbf{Language and Cognitive Development}: 
        Parent and child play a toy-hiding game to foster engagement.}
    \end{subfigure}
    \hfill
    \begin{subfigure}[t]{0.31\textwidth}
        \includegraphics[width=\linewidth]{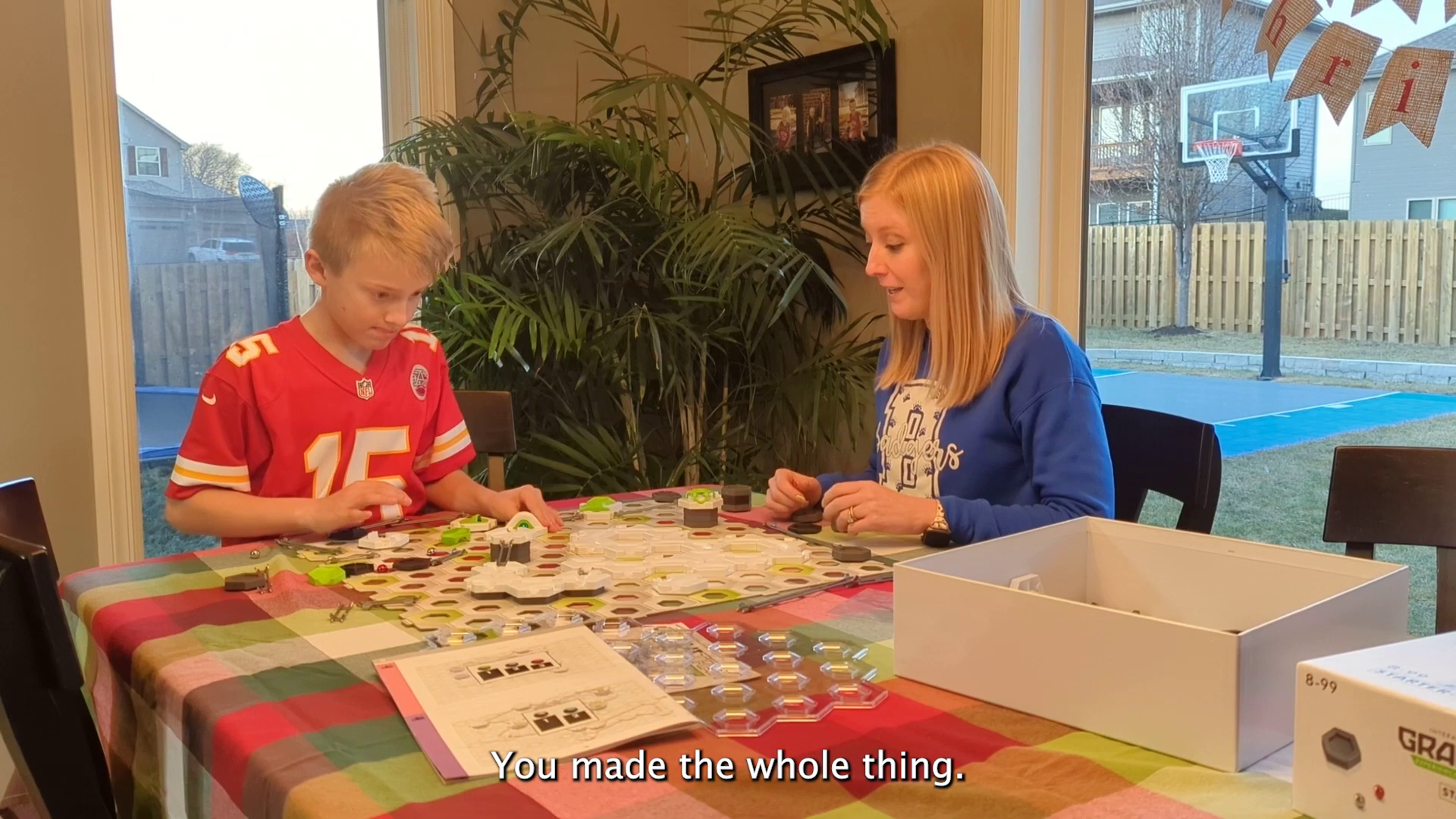}
        \caption{\textbf{Daily Life Skills and Interaction}: 
        Parent and child build a marble run together, encouraging planning.}
    \end{subfigure}
    \caption{Examples from three categories in our dataset: behavioural guidance, language development, and daily life interaction.}
    \label{fig:video_examples}
\end{figure}

Examples from each category are illustrated in Figure~\ref{fig:video_examples}.

%% file: sections/4_slp_interview.tex
\section{UNDERSTANDING SLP OBSERVATION AND JUDGEMENT OF JOINT ATTENTION}
\label{sec:slp_interview}

To address RQ1, we conducted a two-phase study with the same three experienced SLPs: an interview study followed by a annotation study. In the interview study, we carried out semi-structured interviews to elicit how each SLP defines and assesses joint attention in their clinical practice—what behavioural cues they prioritise, and how they distinguish between strong, moderate, and poor instances.

In the subsequent annotation study, each SLP independently reviewed a curated set of parent–child interaction videos. They annotated segments they considered strong or poor in joint attention and provided verbal justifications for each judgement. Segments not explicitly labelled were treated as moderate by default. This process yielded paired annotations and explanatory reasoning, enabling us to extract the key behavioural cues and evaluative logic underlying expert decision-making.

We intentionally separated the interview and annotation studies to ensure that both the conceptual framing and observational criteria emerged directly from the experts themselves. Rather than imposing researcher-defined categories, our goal was to capture the full spectrum of expert reasoning—from high-level definitions to moment-to-moment behavioural observation and judgements—as expressed and enacted by the SLPs.

\subsection{Participants}
Our three expert SLPs were all female, aged between 30 and 59 years (see Table~\ref{tab:slpprofiles}). All have extensive clinical experience working directly with children, particularly those on the autism spectrum, and have received training in a wide range of evidence-based intervention programmes, including Hanen’s \textit{It Takes Two to Talk}~\cite{pepper2004talk}, \textit{More Than Words} and \textit{DIR Floortime}~\cite{dir_floortime}.

Their clinical practice spans public and private sectors and includes diverse settings such as early intervention centres, preschools, home-based care, and multidisciplinary teams. Notably, the SLPs have worked in different countries across multiple continents, contributing to a culturally informed understanding of parent–child interaction.

Given their specialised training and long-standing experience—particularly with children on the autism spectrum, who often experience joint attention difficulties—the SLPs were well-qualified to identify and evaluate strong and poor instances of joint attention in naturalistic contexts.

\begin{table*}[ht]
    \centering
    \begin{tabularx}{\textwidth}{lllXX}
        \toprule
        \textbf{ID} & \textbf{Age} & \textbf{Gender} & \textbf{Qualifications} & \textbf{Experience and Training} \\
        \midrule
        SLP1 & 30–34 & Female & BSc in Speech Pathology; Certified in Hanen (\textit{It Takes Two to Talk}, \textit{More Than Words}); \textit{DIR Floortime} & 7 years with experience in early childhood communication support \\
        
        SLP2 & 30–34 & Female & Bachelor of Speech Pathology; Certified in Hanen (\textit{It Takes Two to Talk}, \textit{More Than Words}); \textit{DIR Floortime}; \textit{PECS}; \textit{PROMPT}; \textit{Social Thinking}; \textit{TalkTools}; \textit{Key Word Sign Australia} & 9 years of paediatric experience across multiple early communication programmes, including sensory processing and oral placement therapy approaches \\
        
        SLP3 & 55–59 & Female & Bachelor of Speech Pathology; Master’s in Communication Disorders; Certified in Hanen (\textit{It Takes Two to Talk}, \textit{More Than Words}); \textit{DIR Floortime} & Over 30 years of clinical experience, including 22 years specialising in paediatric therapy; extensive training in Hanen, DIR Floortime, and international practice standards \\
        \bottomrule
    \end{tabularx}
    \caption{Profiles of our three expert SLPs who participated in the annotation and interview study.}
    \label{tab:slpprofiles}
\end{table*}

\subsection{Method}

To investigate how expert SLPs identify and evaluate joint attention in real-world interactions (RQ1), we conducted a two-part study: an interview study followed by a annotation study. This design allowed us to examine how SLPs conceptualise joint attention in theory and how they apply these concepts in practice—revealing both consistencies and divergences in expert reasoning. To focus our analysis and ensure task feasibility, we limited annotations to the child’s joint attention behaviour, which is often a primary concern in early developmental assessment.

In the interview study, we conducted semi-structured interviews to elicit each SLP’s working definition of joint attention. We asked them to describe what constitutes strong, moderate, or poor joint attention in their professional judgment, what behavioural cues they typically attend to, and how contextual factors may influence their evaluations. These interviews established the criteria SLPs rely on when reasoning about interaction quality.

\begin{figure}[h]
    \centering
    \includegraphics[width=\linewidth]{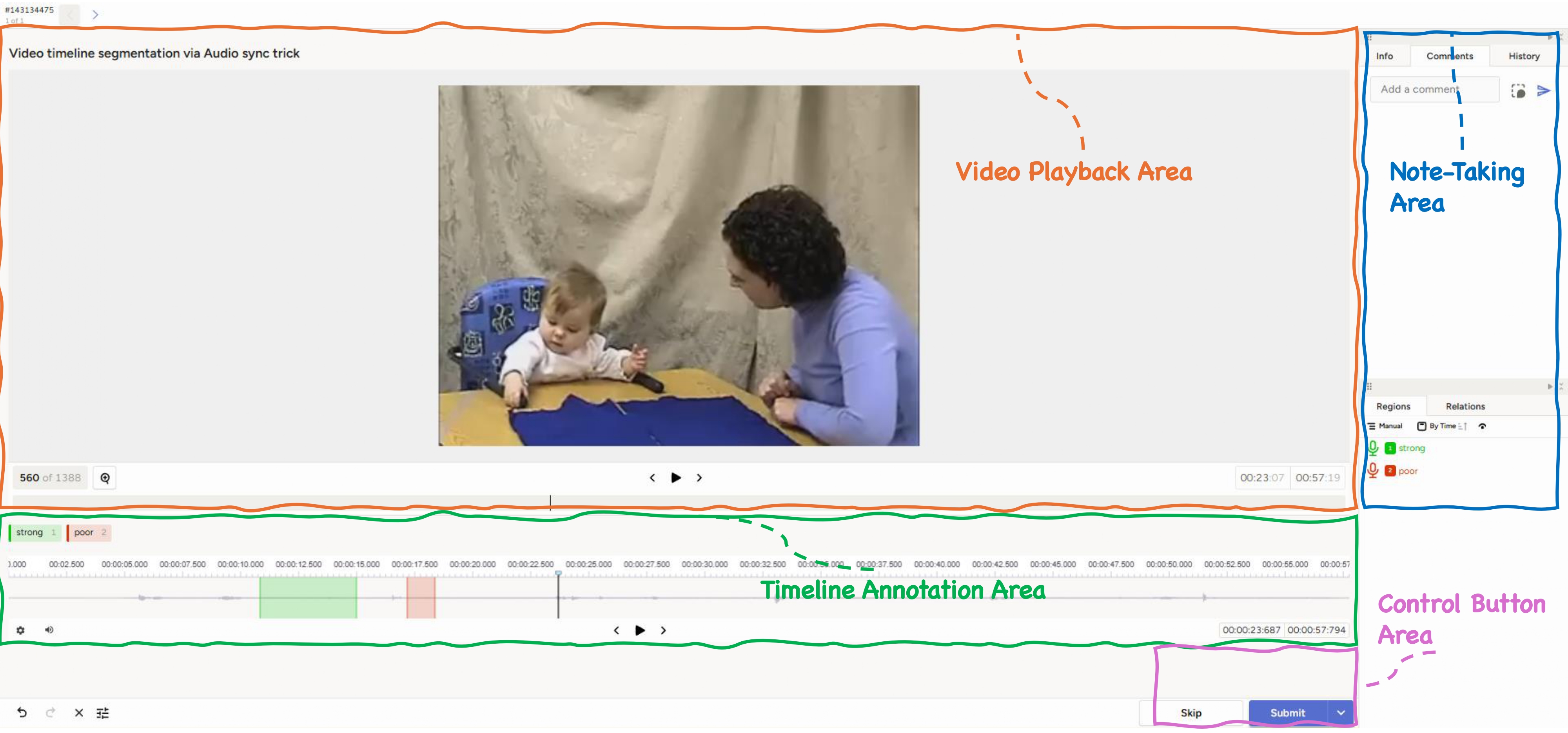}
    \caption{
Our video annotation tool supports SLPs’ judgement process with four main components: 
\textcolor{orange}{Video Playback Area} for watching, pausing, and replaying parent–child interactions; 
\textcolor{darkgreen}{Timeline Annotation Area} for selecting and labelling segments as \textit{strong} or \textit{poor} joint attention; 
\textcolor{navyblue}{Note-Taking Area} for recording justifications or observations; and 
\textcolor{violet}{Control Button Area} for task submission and navigation.
}
    \label{fig:label-tool}
\end{figure}

In the annotation study, each SLP independently reviewed a curated video dataset of 26 parent–child interactions. For each video, they were asked to annotate segments they considered to be clear examples of strong or poor joint attention, with all unlabelled segments treated as moderate by default. Each annotation was accompanied by a short explanation justifying their decision. These explanations provided insight into how SLPs integrate multimodal cues to make holistic judgments during real-time observation. We used a custom video annotation tool to support SLPs’ judgement process (see Figure~\ref{fig:label-tool}).

Together, the interview and annotation studies yielded a dataset of labelled segments and expert rationales, which we used to extract common behavioural cues and assess the consistency between stated criteria and applied judgments. This formed the foundation for developing an MLLM-based system capable of aligning with and simulating expert reasoning. Understanding how SLPs articulate and apply their evaluation criteria is essential for translating these insights into effective model prompts and task designs.

Participants were paid \$10 and \$30 upon completion of the interview and jugdgment study respectively.
All study procedures followed ethical guidelines approved by our institution’s ethics review board.

\subsection{Interview Study Results: SLPs’ Conceptual Understanding of Joint Attention}

Before asking the SLPs to annotate video segments, we first conducted semi-structured interviews to understand how each expert conceptualises joint attention. We asked them to define what joint attention means in their clinical judgment, how they differentiate between strong and poor instances, and whether they expect developmental differences in how joint attention is expressed across child age groups. We summarised key insights from the interviews by grouping similar responses based on how the SLPs described and distinguished different forms of joint attention.
Table~\ref{tab:slp-concepts} summarises their responses.

\begin{table*}[ht]
\centering
\begin{tabularx}{\textwidth}{lXXXX}\toprule
\textbf{ID} & \textbf{How They Define the Child's Joint Attention} & \textbf{What Counts as Strong vs. Poor} & \textbf{How Developmental Stage Affects Expression} & \textbf{How Joint Attention Appears During Play} \\
\midrule
SLP1 & 
Joint attention is the child’s ability to share focus and intent with another person, not just look at the same thing. It's about social connection. & 
Strong: back-and-forth referencing, shared interest \newline
Poor: no acknowledgment of other person’s presence & 
Yes — younger children often rely more on gesture, posture, or affect due to limited language and attention control. For older children, SLP1 expects clearer behavioural signals and social reciprocity. & 
Joint attention fluctuates rapidly during play, especially in younger children. Distraction is expected and not inherently problematic. What matters is the overall balance—occasional poor moments are acceptable as long as strong episodes also occur. \\
\midrule
SLP2 & 
Joint attention is when a child follows or initiates shared gaze toward an object or action, and shows awareness that the other person is also attending. & 
Strong: gaze shifts, alternating eye contact \newline
Poor: no gaze response, ignores joint bids & 
Yes — early on, gaze behaviour may be less consistent. By age 3, SLP2 expects reliable referential gaze, as cognitive and visual coordination typically improve with development. & 
Joint attention during play is highly variable and can shift within seconds. It’s normal for both strong and poor moments to co-occur within a short time window, such as within the same minute. \\
\midrule
SLP3 & 
Joint attention is a shared emotional and cognitive moment, often involving mutual engagement and affective alignment. It’s not just about behaviour but about intention. & 
Strong: emotional reciprocity, coordinated interaction \newline
Poor: child is disengaged despite cues & 
Yes — SLP3 accounts for language, motor, and emotional maturity. For younger children, subtle affective signals may be enough. For preschoolers, SLP3 looks for intentional behaviours like pointing or verbal expression. & 
SLP3 believes it is unrealistic for a child to maintain strong joint attention throughout an entire play session. A child who always appears highly attentive—for example, staring continuously at the parent—would seem unusual. SLP3 described joint attention during play as something that “rises and falls like a temperature bar,” with natural shifts between strong and poor moments. What matters is the overall balance across the interaction, not perfection.\\
\bottomrule
\end{tabularx}
\caption{SLPs’ conceptual definitions of joint attention, their evaluative criteria, and their expectations during play across developmental stages.}
\label{tab:slp-concepts}
\end{table*}

These interviews revealed that while all three SLPs shared a common understanding of joint attention as a socially coordinated process, they differed in how they conceptualised and prioritised specific aspects of it. SLP1 emphasised the coordination of multiple communicative behaviours, and vocalisation—within dynamic, reciprocal interactions between child and parent. SLP2 focused more narrowly on whether the child performed clear gaze shifts between the adult and the shared object, treating visual referencing as the primary indicator of joint attention. In contrast, SLP3 viewed engagement as the core marker, paying particular attention to whether the child demonstrated emotional or attentional alignment with the adult, even in the absence of explicit cues like gaze or pointing.

All three experts agreed that joint attention can manifest differently across developmental stages, influenced by a child’s evolving language, motor, and attention capacities. For younger children—particularly those under age two—they allowed for broader interpretations, including reliance on gestures, posture, or affective responses. For older children, they expected more deliberate and referential signals, such as consistent gaze shifts, pointing, or verbal referencing, reflecting increased communicative intention and control.

Building on these conceptual insights, we next asked each SLP to annotate segments from our curated video dataset. This annotation study allowed us to observe how their stated definitions translated into actual evaluations of strong and poor joint attention in real-world parent–child interactions.

\subsection{Annotation Study Results: SLPs’ Observational Judgements of Joint Attention}

During the annotation process, we observed that all three SLPs consistently relied on three core behavioural dimensions when evaluating joint attention. These dimensions—\textbf{\textit{gaze}}, \textbf{\textit{action}}, and \textbf{\textit{vocalisation}}—served as the primary basis for assessing both the child’s engagement and the parent's responsiveness. These cues were also frequently referenced in the SLPs’ verbal justifications, highlighting their prominence in expert reasoning. Below, we describe how each dimension was used in practice:

\begin{itemize}
    \item \textbf{\textit{Gaze}} — Refers to where the child is looking, such as toward the parent's face, hands, a shared object, or away from the interaction. Gaze cues are critical for interpreting visual attention and shared focus.

    \item \textbf{\textit{Action}} — Encompasses the child's physical behaviours, including pointing, reaching, standing up, dragging objects, or attempting to disengage. These actions signal communicative intent and social participation.

    \item \textbf{\textit{Vocalisation}} — Includes all forms of vocal output, from babbling and laughter to verbal responses. Vocal cues provide insight into the child’s attempt to share attention or respond within the interaction.
\end{itemize}

This triadic lens emerged organically across SLPs and provided a shared framework for interpreting joint attention episodes in a structured yet flexible manner.

To facilitate structured analysis of expert annotations, we divided each video into uniform 5-second segments. While SLPs initially labelled joint attention quality by selecting start and end timestamps based on their judgement, we mapped these annotations onto the fixed segments, assigning each 5-second unit a corresponding label where overlap occurred. This mapping enabled us to translate variable-length judgements into a standardised format, making it easier to compare agreement across experts and analyse annotation patterns consistently across the dataset. Given that the activities—such as guided play or task-based exchanges—are generally short and focused, 5-second segments provided a practical and interpretable unit of analysis.

Table~\ref{tab:slp-distribution} summarises the distribution of joint attention labels assigned by each SLP, as well as the aggregated results based on majority agreement (i.e., at least two SLPs assigning the same label). Across all three experts, the majority of segments were labelled as \textit{Moderate}, suggesting that many observed behaviours fell into an intermediate range that did not clearly indicate strong or poor joint attention. Notably, \textit{Poor} labels were used relatively sparingly by SLP1 and only slightly more frequently by SLP2 and SLP3—reflecting a general reluctance to assign negative evaluations unless disengagement was highly evident.

The combined agreement row reveals that only 22.1\% of segments were jointly considered \textit{Strong}, and just 1.6\% were jointly labelled as \textit{Poor}, while over three-quarters (76.3\%) were consistently judged as \textit{Moderate}. This distribution highlights the interpretive nature of joint attention assessment and the need for careful alignment when designing systems to simulate expert reasoning.

\begin{table*}[ht]
\centering
\caption{Distribution of Joint Attention Labels across SLPs and Combined Agreement ($\geq$2 SLPs)}
\begin{tabular}{lccccccc}
\toprule
\textbf{ID} & \textbf{Strong} & \textbf{Moderate} & \textbf{Poor} & \textbf{Strong (\%)} & \textbf{Moderate (\%)} & \textbf{Poor (\%)} \\
\midrule
SLP1 & 155 & 472 & 11 & 24.3 & 74.0 & 1.7 \\
SLP2 & 195 & 392 & 51 & 30.6 & 61.4 & 8.0 \\
SLP3 & 150 & 437 & 51 & 23.5 & 68.5 & 8.0 \\
\midrule
\textbf{Combined ($\geq$2 agree)} & \textbf{136} & \textbf{469} & \textbf{10} & \textbf{22.1} & \textbf{76.3} & \textbf{1.6} \\
\bottomrule
\end{tabular}
\label{tab:slp-distribution}
\end{table*}

\subsubsection{Illustrative Examples of Consensus Among SLPs}

While expert disagreement was common across many ambiguous segments, a smaller set of segments achieved full agreement among all three SLPs. These moments of consensus reveal the behavioural patterns that experts collectively interpret as clear evidence for strong, moderate, or poor joint attention. In general, jointly labelled \textit{Strong} segments featured multiple forms of engagement—such as mutual gaze, verbal responsiveness, and shared task focus—while \textit{Poor} segments were marked by disengagement, lack of social referencing, or solitary behaviour. Meanwhile, segments labelled \textit{Moderate} often involved task participation without clear signs of shared social coordination. 
The following examples help clarify the implicit thresholds that delineate high-confidence assessments in expert judgement:

\begin{itemize}
    \item \textbf{Video A Segment 012}: In this segment, the parent pointed toward a set of toys and instructed the child to listen. The child held a toy and expressed a desire to build. Although the child’s gaze remained focused on the object, their verbal response and engagement with the task were strong. All three SLPs rated the segment as \textit{Strong}, citing meaningful participation and clear communicative intent.
    \item \textbf{Video B Segment 032}: The parent placed her hands near some coins and gave verbal instructions. The child echoed the parent’s words while briefly looking at her. This multimodal reciprocity—mutual gaze and verbal alignment—led all experts to label the segment as \textit{Strong}, highlighting it as a clear example of shared attention.
    \item \textbf{Video C Segment 103}: In this segment, the parent picked up train tracks, and the child silently connected the pieces. Although neither party spoke, both were visually focused on the same task and remained jointly engaged. The absence of verbal exchange or referencing gestures placed this segment in a middle zone. All three SLPs agreed on a \textit{Moderate} label, recognising the co-attention but noting the limited social intent.
    \item \textbf{Video D Segment 033}: In this interaction, the parent made an encouraging comment while facing the child. However, the child remained fixated on the toys and avoided looking at the parent. There was no observable attempt to share attention or respond socially. All three SLPs labelled this segment as \textit{Poor}, agreeing that the absence of gaze and referencing indicated disengagement from the social interaction.
\end{itemize}

These consensus segments demonstrate how, despite individual stylistic differences, experts converge when multiple cues co-occur (for \textit{Strong}), are present but non-social (for \textit{Moderate}), or are entirely absent (for \textit{Poor}). Such segments serve as useful anchors when training or evaluating AI models that aim to simulate SLP-like assessments.

\subsubsection{Illustrative Examples of Diverging SLP Judgements}

Although all three SLPs shared a general understanding of joint attention as a coordinated social process, their specific criteria and interpretive emphasis varied. These differences became especially evident when they encountered ambiguous or intermediate segments.

\textbf{SLP1} adopted a balanced and multimodal approach. While gaze was still considered, it was treated as one of several contributing cues—alongside gestures, verbal referencing, emotional affect, and body orientation. SLP2 often acknowledged child-led play and nonverbal forms of engagement, showing flexibility in how joint attention could be expressed. Segments with inconsistent gaze but strong affective presence were still given high ratings when the child’s overall behaviour reflected social coordination.

\textbf{SLP2}, in contrast, placed strong emphasis on visual engagement, particularly gaze alternation between the adult and shared object. Eye contact and sustained visual referencing were seen as essential for identifying strong joint attention. In the absence of clear gaze cues, even segments involving verbal speech or physical interaction were often rated as moderate or poor. For example, SLP1 rated several segments as poor when the child appeared engaged but did not meet the baseline requirement of mutual gaze.

\textbf{SLP3} took a functional and context-sensitive perspective, often evaluating segments based on the child’s communicative intent rather than adherence to typical behavioural markers. Rather than expecting conventional cues like pointing or verbalisation, SLP3 recognised alternative forms of participation—such as pushing away the parent’s hand or physically repositioning themselves—as valid signs of engagement. This expert was less concerned with surface-level eye contact and more attentive to whether the child demonstrated awareness and responsiveness in their own way.

To illustrate these differences, we highlight several segments where the three SLPs disagreed in their assessments. These examples underscore the unique interpretive lenses each expert applied:

\begin{itemize}
    \item \textbf{Video E Segment 003}: In this segment, the child sat on the floor playing with a toy airplane and said, “I want to go to the Bahamas. Fly there now.” While there was no gaze toward the parent, the child’s speech was contextually rich. SLP1, recognising the imaginative verbal output and engagement, rated it as \textit{Moderate}. SLP2 labelled the segment as \textit{Poor} due to the lack of visual coordination. SLP3 rated it \textit{Strong}, interpreting the child’s verbal and motor actions as clear evidence of communicative intent.
    \item \textbf{Video F Segment 004}: The child alternated gaze between a toy and the parent’s hands while reaching for the toy, but did not speak. SLP1 gave a \textit{Moderate} rating, due to the absence of vocal engagement. SLP2 labelled this as \textit{Strong}, citing the clear visual coordination. SLP3 judged it as \textit{Poor}, arguing that the interaction lacked reciprocal cues or intentional signalling.
    \item \textbf{Video G Segment 017}: In this segment, the child pointed to a toy lion without speaking. SLP1 rated it \textit{Moderate}, perhaps noting the absence of emotional or verbal expression. SLP2 rated the interaction as \textit{Strong}, identifying the combination of gaze and pointing as sufficient for joint attention. SLP3 gave it a \textit{Poor}, interpreting the gesture as mechanical rather than socially directed.
\end{itemize}

Together, these examples demonstrate how gaze, vocalisation, and action were weighted differently by each expert, and how their background and theoretical orientation shaped their interpretation

%% file: sections/5_mllm_system.tex
\section{ALIGNING MLLMS WITH SLP: BEHAVIOUR OBSERVATION-TO-JUDGEMENT}
\label{sec:mllm_system}

Building on our findings from the interview and annotation studies, we sought to translate SLPs’ reasoning processes into a computational framework. Specifically, we developed a two-stage system that enables MLLMs to simulate how SLPs observe and judge joint attention. The first stage focuses on structured behavioural observation—using the three expert-derived dimensions of \textit{gaze}, \textit{action}, and \textit{vocalisation} to prompt the MLLM to describe what is happening in each video segment. The second stage then assesses whether these structured descriptions improve the MLLM’s ability to judge the quality of joint attention. We evaluate this by comparing zero-shot and few-shot prompting strategies, and contrasting reasoning-based and non-reasoning model variants. Through this pipeline, we examine how well MLLMs can align with expert criteria in both perception and judgement.

\subsection{Stage 1: Behaviour Description through Expert-Aligned Prompting}

\subsubsection{Method}

With the rapid advancement of MLLMs, their application to human behaviour analysis presents a promising frontier. In this study, we adopt \texttt{Gemini 2.5 Pro}\footnote{Gemini 2.5 Pro: \url{https://blog.google/products/gemini/gemini-2-5-pro-updates/}}, a state-of-the-art MLLM that supports direct video input. Gemini has demonstrated strong performance on \textit{Video-MME}~\cite{fu2024video}, a recent benchmark evaluating video understanding across diverse domains and modalities, outperforming other commercial and open-source models in tasks requiring multimodal reasoning over time.
However, prior research~\cite{shi2025towards} shows that directly prompting MLLMs to identify key interaction segments—-especially for complex, socially embedded tasks like joint attention—often fails. Models continue to struggle with long video understanding, temporal grounding, and fine-grained cues such as gaze interpretation.

To address these limitations, we build on insights from our SLP interviews, which revealed two important features of joint attention evaluation: (1) key behavioural changes are rapid and often context-independent, and (2) expert assessments consistently rely on three observable cues—\textbf{gaze}, \textbf{action}, and \textbf{vocalisation} (see Table~\ref{tab:slp-concepts}).

Guided by these principles, we divided each video into uniform 5-second segments and designed a structured prompting scheme to elicit behavioural summaries from the MLLM. For each segment, the model was asked to describe both the parent and the child’s behaviour along the three core dimensions. To encourage accurate, grounded behavioural descriptions, we designed a zero-shot prompt that guides the model to produce short, factual observations in natural language. By enforcing a consistent subject–verb–object structure, the prompt reflects the interpretive style of human annotators and reduces the likelihood of hallucinated or overly abstract outputs. The full prompt is shown below:

\noindent\textbf{\textit{Structured Behaviour Description Prompt (Gaze–Action–Vocalisation)}}:
\begin{quote}\ttfamily
You are watching a video of a parent interacting with a child.

For each participant (parent and child), describe their behaviour in three parts:

1. Gaze: Describe what or whom the person was looking at, using a natural language sentence with a subject–verb–object structure.  
   Examples:  
   - The child looked at the parent’s face.  
   - The parent shifted gaze between the child and the toy.  
   - The child stared at the blocks.  
   - The parent looked away for a moment.

2. Action: Describe what the person physically did, using a short natural language sentence with a subject–verb–object structure.  
   Examples:  
   - The parent pointed at the red ball.  
   - The child reached towards the puzzle pieces.  
   - The parent lifted the toy truck.  
   - The child clapped their hands.

3. Vocalisation: Transcribe or paraphrase what the person said, or describe any vocalisations using a short sentence with a subject–verb–object structure.  
   Write 'None' if there were no vocalisations.
\end{quote}

\subsubsection{Result}
We ran the full set of behaviour description prompts on an Ubuntu 22.04 LTS server via the official \texttt{Gemini 2.5 Pro} API, collecting MLLM-generated outputs for all 5-second video segments in our dataset.

To evaluate the MLLM’s behavioural description performance, we compared its outputs with reference annotations created by our research team. These reference annotations were constructed in two stages, following the same gaze–action–vocalisation structure used in the prompt. Importantly, our annotation process was informed by the interview findings with SLPs—particularly their attention to detail and tendency to distinguish subtle variations in gaze direction, gesture context, and vocalisation specificity.

For each segment, we began by manually reviewing the MLLM-generated outputs. In the first stage, we corrected instances where the generated description contradicted the observed behaviour—for example, changing “the child looked at the parent” to “the child looked down at the table” if no eye contact was actually made. In the second stage, we refined underspecified descriptions by aligning them with SLP-like granularity—for instance, revising “the child looked at the parent” to “the child looked at the parent’s hand” when the visual target was more specific. These two-step corrections ensured that reference labels reflected both factual accuracy and expert-informed attentional focus.

Based on this corrected reference set, we computed segment-level accuracy scores for each of the three behavioural fields. As shown in Table~\ref{fig:mllm}, the MLLM performed best in the \textit{action} category, with a mean accuracy of 0.88, followed by \textit{vocalisation} (0.87) and \textit{gaze} (0.86). All three fields achieved perfect accuracy in at least one video, but \textit{gaze} also had the lowest minimum, highlighting its difficulty for the model.

\begin{table}[h]
\centering
\caption{Accuracy statistics by behavioural field}
\begin{tabular}{lcccc}
\toprule
\textbf{Field} & \textbf{Mean} & \textbf{Median} & \textbf{Max} & \textbf{Min} \\
\midrule
Action        & 0.8774        & 0.9464          & 1.0000       & 0.6250       \\
Vocalisation  & 0.8708        & 0.9259          & 1.0000       & 0.5000       \\
Gaze          & 0.8556        & 0.8750          & 1.0000       & 0.5000       \\
\bottomrule
\end{tabular}
\label{fig:mllm}
\end{table}

To better understand model failure cases, we conducted a qualitative review of low-performing videos across the three behavioural fields. Several recurring issues emerged:

\begin{itemize}
  \item \textbf{Speech role confusion} was a major factor affecting vocalisation accuracy. For instance, in \textit{Video A} and \textit{Video B}, the model consistently attributed child-like vocalisations to the parent—particularly when the parent mimicked the child’s babbling. Similar confusion was observed in \textit{Video C}, where rapid turn-taking and overlapping speech made speaker attribution unreliable.
  
  \item \textbf{Gaze misinterpretation} often occurred when faces were partially occluded, or when the child looked at non-face targets such as hands or objects. In \textit{Video D}, the child’s gaze was repeatedly marked as “looking at the parent” despite clear visual evidence to the contrary. Likewise, in \textit{Video E}, gaze was mislabelled even when the parent was not in the frame.
  
  \item \textbf{Action detection errors} were more frequent in unstructured scenes. For example, in \textit{Video C}, the child’s aggressive toy-hitting was not recognised, and in \textit{Video F}, a clear tray-passing motion was entirely missed. In \textit{Video G}, task misunderstanding led to mismatched labels (e.g., writing mistaken for eating).
\end{itemize}

These results demonstrate that our structured prompting approach substantially improved the reliability of MLLM-generated behavioural annotations, enabling the model to produce interpretable outputs that aligned with expert-labelled segment judgments in many cases. Compared to Shi et al.~\cite{shi2025towards}, where average behavioural and gaze accuracy was around 62\%, our structured prompting approach led to substantial improvements in both description accuracy and eye-contact interpretation.
While challenges remain—particularly in occluded, ambiguous, or noisy interaction contexts—our findings offer actionable guidance: to support more accurate MLLM observation of social behaviour, future designs should prioritise precise prompt structures, segment-level granularity, and explicit modelling of role-specific cues such as gaze targets and vocal turn-taking.

\subsection{Stage 2: Joint attention judgement through Expert-Aligned Prompting}

\subsubsection{Method}

In the previous stage, we prompted MLLMs to generate structured behavioural descriptions—capturing key cues such as gaze, action, and vocalisation—based on SLP-inspired observation practices. We now extend this pipeline to examine a core question: \textit{can MLLMs simulate SLP-like judgement of joint attention based on these descriptions?}

To evaluate this, we test two dominant families of MLLMs: a non-reasoning model (\texttt{GPT-4o mini}\footnote{GPT-4o mini: \url{https://platform.openai.com/docs/models/gpt-4o-mini}}) and a reasoning-oriented model (\texttt{o4-mini\footnote{o4-mini: \url{https://platform.openai.com/docs/models/o4-mini}}}), both under zero-shot and few-shot prompting conditions. These represent two popular paradigms in current MLLM research—direct classification vs. step-by-step reasoning—and allow us to examine how different forms of prompting affect alignment with expert judgements.
We chose not to use the advanced \texttt{GPT-4o} model due to its significantly higher computational cost. Given that our structured prompting approach decomposes the judgement task into simpler, interpretable components, we aimed to evaluate whether smaller, cost-efficient models could perform reliably. Both \texttt{GPT-4o mini} and \texttt{o4-mini} are lightweight and support fine-tuning, making them well-suited as flexible baselines for future work. Although our current study focuses on prompting only, we selected these models to ensure compatibility with potential fine-tuning and other adaptation methods in follow-up studies.

Drawing from our earlier interviews, we observed that SLPs typically arrive at a judgement through three steps: observing behavioural cues, reasoning about social coordination, and mapping these to categorical Judgements (e.g., strong, moderate, poor). We therefore designed the prompt structure to mirror this expert reasoning pipeline.

\textbf{In the zero-shot condition}, both models were given only instructions and structured behaviour input. For \texttt{GPT-4o-mini}, the prompt directly asked for a classification based on the input. For \texttt{o4-mini}, we used a structured three-part template—\texttt{Observation}, \texttt{Reasoning}, and \texttt{Judgement}—to encourage reflective justification before labelling. The full prompt format is shown below:

\noindent\textbf{\textit{Zero-shot prompt (non-reasoning model)}}:

\begin{quote}\ttfamily
You are a speech-language pathologist. Joint attention in a child refers to the ability to share attention with another person—typically the parent—by coordinating behaviours such as actions, vocalisations (e.g., speaking or making sounds), or gaze (looking at shared objects or people).

Please evaluate the quality of the child's joint attention in each segment below based on their behaviours.

Respond using the following format:

Segment 1: [Strong/Moderate/Poor] \\
Segment 2: ...
\end{quote}

In contrast, the non-reasoning model (\texttt{GPT-4o-mini}) was prompted to directly output a label for each segment without justification, following a lightweight, classification-style template:

\noindent\textbf{\textit{Zero-shot prompt (reasoning model})}:

\begin{quote}\ttfamily
You are a speech-language pathologist. Joint attention in a child refers to the ability to share attention with another person—typically the parent—by coordinating behaviours such as actions, vocalisations (e.g., speaking or making sounds), or gaze (looking at shared objects or people).

For each segment below, do the following:

1. Observation: Briefly describe what the parent and child did, said, and looked at. \\
2. Reasoning: Assess whether the child was showing clear, partial, or minimal joint attention, and explain why. \\
3. Judgement: Assign a final label — Strong / Moderate / Poor.
\end{quote}

\textbf{In the few-shot condition}, we augmented each prompt with three reference examples—one for each joint attention category (\textit{strong}, \textit{moderate}, \textit{poor}). Each example was selected from a segment where all three SLPs provided unanimous labels, ensuring high-quality agreement. 

For the non-reasoning model, examples consisted of two fields: \texttt{Observation} and \texttt{Judgement}. This format was designed to provide structural guidance while avoiding interpretive steps. In contrast, the reasoning model was prompted with full triplets—\texttt{Observation}, \texttt{Reasoning}, and \texttt{Judgement}—mirroring the expert judgement process identified in our earlier interviews. The \texttt{Reasoning} field was especially important for eliciting model justification aligned with SLP logic. Each reasoning line was constructed by synthesising verbal explanations given by the three SLPs during annotation. Below, we present the complete set of few-shot examples used for the reasoning model.

\noindent\textbf{\textit{Few-shot examples (reasoning model})}:

\begin{quote}\ttfamily
Segment 1: \\
Observation: \\
Gaze: [Parent] Looked at the toy shelf and occasionally at the child. [Child] Looked at the toys. \\
Action: [Parent] Pointed toward the toy shelf. [Child] Held a toy and remained seated. \\
Vocalisation: [Parent] Instructed the child to listen. [Child] Said they wanted to build. \\
Reasoning: The child showed clear engagement with the task and responded meaningfully, although gaze was primarily on the object. \\
Judgement: Strong

\vspace{0.5em}

Segment 2: \\
Observation: \\
Gaze: [Parent] Looked at the toy pieces. [Child] Focused on the train tracks throughout the segment. \\
Action: [Parent] Picked up a track piece. [Child] Connected the track pieces without hesitation. \\
Vocalisation: [Parent] Did not speak. [Child] Remained silent. \\
Reasoning: Although no verbal exchange occurred, both parties were jointly focused on the same task. Engagement was sustained. \\
Judgement: Moderate

\vspace{0.5em}

Segment 3: \\
Observation: \\
Gaze: [Parent] Looked at the child’s face. [Child] Looked only at the toys. \\
Action: [Parent] Sat facing the child and remained still. [Child] Manipulated the toy without shifting posture. \\
Vocalisation: [Parent] Made an encouraging comment. [Child] Did not respond verbally. \\
Reasoning: The child remained engaged with the activity but showed minimal social referencing or gaze toward the parent. \\
Judgement: Poor
\end{quote}

For the non-reasoning model, the same segment examples were provided without the \texttt{Reasoning} component, containing only the \texttt{Observation} and \texttt{Judgement} fields.

\subsubsection{Model Perspective: Which Prompting Strategy Yields Better Alignment?}

To evaluate how well MLLMs can simulate expert judgement of joint attention, we ran a total of four experimental conditions: \textit{reasoning vs. non-reasoning} models, each under \textit{zero-shot and few-shot} prompting. In each condition, we called the respective model’s API on an Ubuntu 22.04 LTS server, feeding in the full segment-level descriptions for all 26 videos in our dataset.

For the non-reasoning model, the input consisted of behaviour summaries per segment, and the model directly returned a sequence of joint attention judgements for each segment (i.e., \texttt{Strong}, \texttt{Moderate}, or \texttt{Poor}). For the reasoning model, the output included both a structured \texttt{Reasoning} and final \texttt{Judgement} field. Across all runs, we collected model predictions for every 5-second segment in the dataset.

To assess alignment with expert judgement, we compared model-generated labels against segment-level labels from all three SLPs. For each experimental condition, we computed \textbf{precision}, \textbf{recall}, and \textbf{F1 score} per joint attention category—\textit{strong}, \textit{moderate}, and \textit{poor}—relative to each SLP’s annotations. The results were visualised using radar charts to highlight performance differences across conditions.

We use standard definitions for the metrics: \textbf{Precision} measures the proportion of predicted labels that are correct; \textbf{Recall} measures the proportion of SLP labels that were correctly identified by the model; and \textbf{F1 score} is the harmonic mean of precision and recall, representing overall alignment performance.

To visualise model performance across expert perspectives, we generated radar plots for each SLP, comparing the four model configurations (\textit{zero-shot/few-shot} × \textit{reasoning/non-reasoning}) across the three joint attention categories. Each plot shows precision, recall, and F1 score for the \textit{Strong}, \textit{Moderate}, and \textit{Poor} labels. These visualisations highlight how prompting strategy and reasoning capacity affect alignment with different experts.

\begin{figure*}[htbp]
    \centering
    \includegraphics[width=0.95\textwidth]{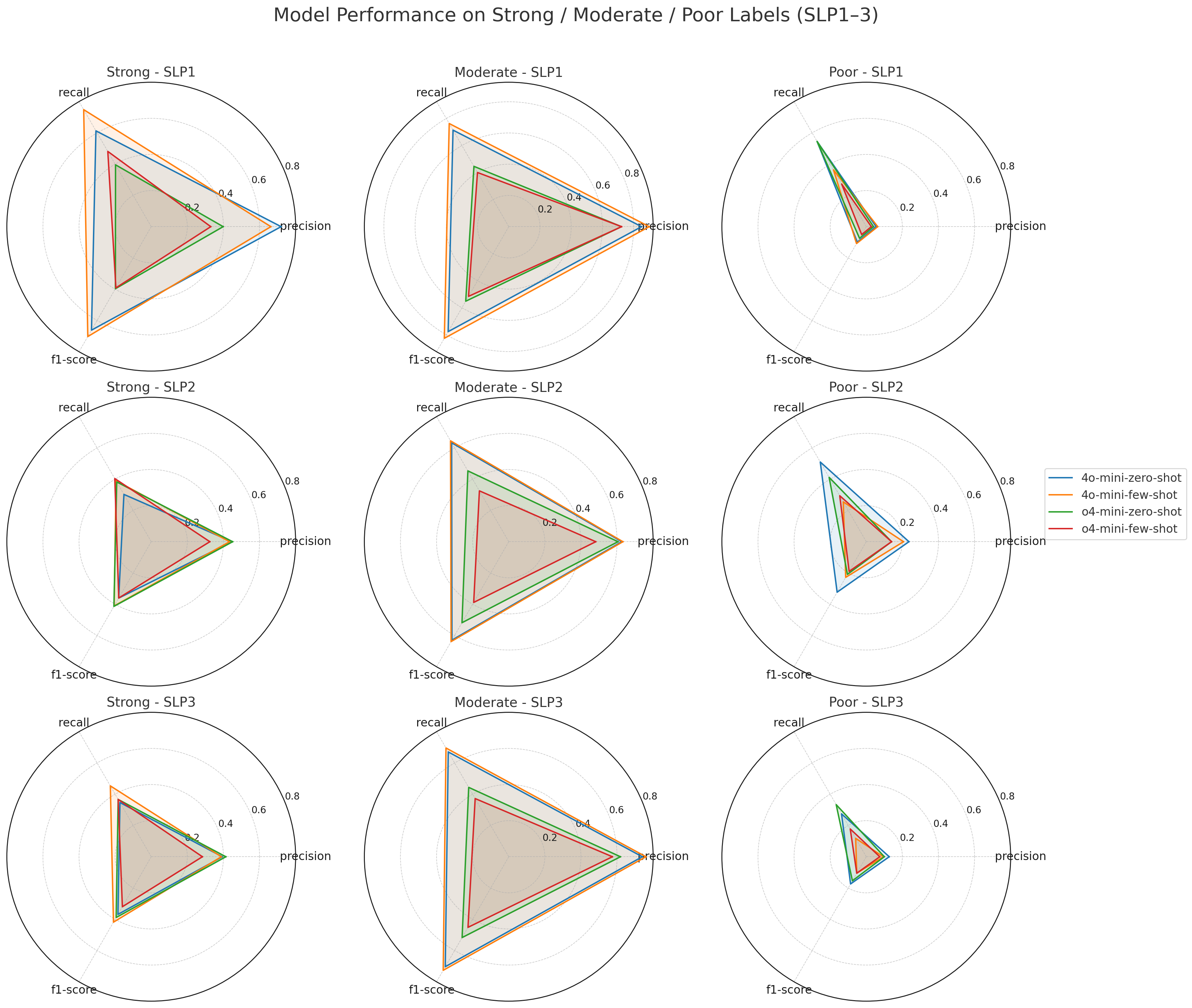}
    \caption{Radar plots showing model performance across all three SLPs. Each subplot compares four model configurations (\textit{zero-shot/few-shot} × \textit{reasoning/non-reasoning}) across the three joint attention labels (\textit{Strong}, \textit{Moderate}, \textit{Poor}), using precision, recall, and F1-score as axes. Each line represents a different model; larger areas indicate better alignment with expert labels.}
    \label{fig:model}
\end{figure*}

Figure~\ref{fig:model} reveals a consistent performance hierarchy across all three SLPs. \texttt{GPT-4o mini (few-shot)} achieved the highest alignment scores, followed by its zero-shot counterpart. Both outperformed the reasoning-based \texttt{o4-mini} variants, with \texttt{o4-mini (few-shot)} yielding the weakest performance across all labels and metrics. This result suggests that few-shot prompting with a non-reasoning model most closely matches the labelling strategies employed by expert SLPs.

This outcome contrasts with common expectations that reasoning-oriented prompting—such as Chain-of-Thought (CoT) generation—should improve interpretability and classification accuracy. However, in our joint attention task, expert judgements were not grounded in explicit reasoning sequences. Instead, SLPs relied on rapid, intuitive interpretation of multimodal behavioural cues—especially gaze, gesture, and vocalisation. Their labels reflected recognisable interaction patterns, not deliberative logic.

Reasoning-based models, particularly in the few-shot condition, frequently produced verbose or misaligned outputs. Without a coherent reasoning structure in the examples to anchor their inferences, these models often defaulted to generalised or overly cautious labelling, leading to an overuse of the \textit{Moderate} category and a failure to discriminate edge cases. This was especially problematic in segments requiring precision—such as borderline \textit{Strong} or \textit{Poor} cases—where vague explanations undermined classification fidelity.

In contrast, non-reasoning models like \texttt{GPT-4o mini} benefited substantially from few-shot prompting. These models did not attempt to explain label decisions but instead mimicked the structural patterns embedded in the examples. Because expert annotations in this context are largely pattern-based and cue-driven, this approach proved more effective. Few-shot prompts served as implicit templates, allowing the model to internalise common combinations of gaze, action, and vocalisation without needing explicit rules.

Overall, our findings indicate that for expert-aligned classification tasks grounded in perceptual judgement rather than structured logic, non-reasoning MLLMs paired with carefully designed few-shot prompts offer superior performance. Conversely, attempts to induce reasoning in such domains may introduce unnecessary complexity, ultimately degrading alignment with human expertise.

\subsection{SLP Perspective: Which Expert Judgement Style Aligns Best with MLLMs?}

Our empirical results show that \texttt{GPT-4o mini (few-shot)} aligned most strongly with SLP1 across all metrics. This includes the highest overall macro precision and F1-score among the three experts. A deeper interpretation reveals several factors that may explain this alignment pattern.

SLP1 applied relatively balanced criteria across gaze, vocalisation, and action. This enabled the model to anchor predictions on well-rounded behavioural patterns. In the \textit{Strong} category, for example, the model achieved its highest alignment scores against SLP1, with a precision of 0.72, recall of 0.61, and F1-score of 0.66. The clarity and consistency in SLP1’s labelling—particularly for multi-modal engagement—provided robust learning signals during few-shot prompting and proved especially effective for reasoning-based models.

SLP2, by contrast, placed disproportionate emphasis on gaze alternation—specifically requiring frequent shifts between parent and object or intentional verbal alignment. This stricter criterion improved the model’s alignment in the \textit{Poor} category, where lack of gaze was a consistent indicator. However, it also lowered recall for \textit{Strong} segments, as the model often misclassified borderline cases with partial gaze cues—even when other modalities indicated strong engagement.

SLP3 adopted a more flexible, engagement-centred lens that privileged subtle contextual cues and affective presence over explicit behavioural markers. This made her labelling style more difficult to simulate using discrete behavioural descriptions. As a result, model alignment with SLP3 was generally lower, particularly in \textit{Strong} and \textit{Poor} categories, where intent was more interpretive than explicit.

Overall, alignment with SLP judgement followed the trend: \textit{Moderate} > \textit{Strong} > \textit{Poor}, driven by both label distribution and task-specific ambiguity. First, \textit{Moderate} segments were most frequent in the dataset, making them statistically more likely to be predicted correctly—especially in the zero-shot condition, where models lacked clearly defined classification criteria. The absence of strict decision boundaries, combined with minimal in-context examples, led models to default to the more neutral Moderate label under uncertainty.

Second, accuracy for \textit{Strong} segments exceeded that of \textit{Poor}, reflecting greater consensus among SLPs and better support from the few-shot examples. In our experimental design, Strong examples were drawn from segments with unanimous expert agreement, offering the model clearer behavioural anchors. In contrast, \textit{Poor} labels were used less frequently by experts and often involved more subjective interpretation—such as disengagement or social withdrawal—making them harder for the model to identify and learn reliably.

%% file: sections/6_design_guide.tex
\section{Design Guidelines for Aligning MLLM Systems with SLP Judgement}

Based on our SLPs interviews and experiments, we outline key implications for designing MLLM-powered systems to support joint attention assessment. These guidelines span system functionality, model configuration, user interaction, and data adaptation.

\subsection{Behaviour-Centric Observation Structure}
SLP judgement relies consistently on three core behavioural cues: \textbf{gaze}, \textbf{action}, and \textbf{vocalisation}. Structuring behavioural analysis around these dimensions allows MLLMs to operate within a bounded, interpretable framework that aligns with expert reasoning. Our findings show that even without additional training, MLLMs can achieve over 85\% accuracy in these fields using structured, zero-shot prompts.

To accommodate nuanced cases—especially in neurodiverse populations—future systems may incorporate a fourth dimension: \textbf{engagement}. This reflects sustained interaction, emotional presence, or participation that does not always manifest through gaze or speech. While MLLMs currently struggle to infer such abstract signals from limited frames, caregiver interaction (e.g., confirming child interest) or affective sensing (e.g., posture, movement continuity) could augment this gap. For applications targeting parents, guided input tools (e.g., swipe-to-confirm gaze, voice-tagging) may serve as low-friction ways to supplement model observations.

\subsection{Few-Shot Prompting with Retrieval-Augmented Grounding}
Our results indicate that providing a small number of expert-labelled examples significantly boosts MLLM alignment. Non-reasoning models (e.g., GPT-4o mini) prompted with just three few-shot examples achieved robust precision and recall on both "Moderate" and "Strong" segments. However, agreement with "Poor" labels remained challenging due to limited training signal and ambiguity.

Future implementations should consider \textbf{retrieval-augmented prompting}, where a system selects a tailored set of examples based on interaction type, developmental age, or caregiver input. This mechanism can dynamically fetch age-appropriate examples or common misalignment cases (e.g., non-verbal engagement), providing personalised grounding for MLLM evaluation. This approach would improve generalisability without requiring fine-tuning.

\subsection{Avoid Over-Structured Reasoning Formats}
Contrary to expectations, reasoning-style models using Observation--Reasoning--Judgement templates underperformed in our experiments. This suggests that expert decision-making in joint attention is often \textbf{non-explanatory}, relying on intuitive, pattern-based recognition of behavioural cues.

In contrast to domains like mathematics or logic puzzles where step-by-step reasoning enhances accuracy, parent–child interaction is inherently behavioural. Joint attention judgements rely more on the perception and interpretation of visible social cues than on verbal justification. Encouraging the model to simulate reasoning through template-based prompts did not improve interpretability; in fact, it reduced precision due to verbose, off-topic outputs and label confusion. In tasks like joint attention classification, models benefit more from \textbf{copyable, structural templates} than from logic-based deduction. Designers should prioritise example-driven prompting over synthetic justification chains.

\subsection{Developmental and Neurodiversity Adaptation}
Joint attention is not static—it evolves with age, context, and individual differences. Our SLP interviews confirmed that expectations for joint attention vary based on the child’s developmental stage. For children under two, gaze cues may be weak, and body orientation or affective tone become primary indicators. Conversely, by age four, intentional pointing and verbal reference are expected.

Systems should allow for \textbf{adaptive weighting of cues} based on age or developmental profile. For example, in children with autism, gaze may be inconsistent, but joint attention can still manifest through action (e.g., bringing objects) or vocalisation (e.g., echolalia). MLLMs can benefit from custom thresholds or cue prioritisation models to avoid under-classification of valid engagement in atypical cases.

This adaptation can be implemented through modular classifiers or condition-specific prompting libraries. For end-users, interface features such as selecting age range or developmental concerns (e.g., "early-stage non-verbal") can guide the system's output tone and decision tolerance.

Together, these design principles enable the development of AI systems that are sensitive to expert strategies, supportive of parental learning, and scalable across contexts.

%% file: sections/7_discussion.tex
\section{Discussion}

\subsection{The Need for Diverse SLP Standards and Annotated Video Data}

While our study demonstrates promising alignment between MLLMs and SLPs in evaluating joint attention, the generalisability of these results remains constrained by the limited scope of our dataset. Our analysis was based on a small number of expert raters and a modest corpus of curated online videos, most of which featured neurotypical children in relatively structured play settings.

To advance the broader utility of our framework, future work must expand in two directions. First, incorporating annotations from a more diverse pool of SLPs—including those with different training backgrounds, clinical philosophies, and geographic or cultural contexts—could surface richer patterns of expert reasoning and provide more robust grounding for model alignment. Second, the dataset itself must be diversified to reflect a wider spectrum of developmental trajectories. In particular, including videos of children with atypical development (e.g., those on the autism spectrum, with ADHD, or other communication differences) is essential for modelling the variability of joint attention cues in real-world clinical scenarios. For instance, Autistic children often exhibit reduced gaze behaviour but compensate through vocalisation or gestural engagement—patterns that are not well captured in current datasets or MLLM prompts.

Moreover, increasing the quantity and quality of labelled data opens opportunities for direct fine-tuning of MLLMs, rather than relying solely on prompting. A larger and more representative set of parent–child interaction videos—with clearly demarcated and consensus-annotated instances of “Strong”, “Moderate”, and “Poor” joint attention—would allow for supervised learning methods that go beyond in-context simulation. Such alignment datasets could establish new benchmarks for human-aligned behavioural modelling, enabling model evaluation that is both more transparent and clinically relevant.

\subsection{Exploring Parent-AI Interaction in Joint Attention Assessment}

Beyond expert alignment, our findings point to an emerging opportunity to incorporate parents themselves into the assessment loop—particularly in the context of supporting AI-assisted reflection on their everyday interactions. In the early phases of our study, we observed that parents often contributed contextual interpretation when reviewing videos, such as noting when the child was tired, resistant, or unusually distracted. These parent observations, while not clinical in nature, can offer grounding signals for ambiguous segments where expert-only annotations may overlook context-specific meaning.

Future AI-assisted systems for joint attention assessment should explore parent–AI co-annotation workflows. For example, since MLLMs currently struggle with precise gaze estimation—especially when faces are partially occluded or turned away—parents could be invited to verify or annotate where the child was looking during key moments. This hybrid approach may improve the interpretability and accuracy of model outputs, especially in settings where high-stakes judgments are made using limited or noisy input.

Furthermore, we argue that developmental feedback systems must eventually adapt to individual child profiles. A general-purpose framework may suffice for neurotypical children, but children with atypical communication profiles require more flexible and personalised analysis pipelines. These could be informed not only by expert-derived cue structures (e.g., gaze–action–vocalisation), but also by parental observations of how their child typically initiates, sustains, or avoids joint attention. In this sense, parents are not just passive recipients of AI feedback but potential collaborators in helping calibrate behavioural understanding.

\subsection{Toward a Scalable, Personalised, and Human-Centred Framework}

Our study offers an initial step toward a scalable and interpretable system for modelling joint attention from video data. While our methods rely on relatively simple prompting and few-shot examples, the results show that MLLMs are capable of aligning with expert standards to a meaningful extent—especially when behaviour is structured and cues are observable. That said, the variability in both expert criteria and child behaviour underscores the need for a more modular and adaptive system. One promising direction involves combining prompt-based learning with retrieval-augmented generation, allowing systems to reference diverse, expert-validated examples in real time. For instance, families could upload annotated clips of their child’s interaction history, which the model can retrieve and align with expert baselines when making new judgements.

We envision future developmental AI tools that do not simply mimic SLP decisions, but dynamically integrate model prediction, expert logic, and parent insight. To achieve this, future research should address three key challenges: (1) expanding datasets to include non-typical developmental trajectories and more diverse cultural expressions of attention; (2) capturing the evolving, age-dependent manifestations of joint attention across infancy and early childhood; and (3) enabling parents to participate meaningfully in behaviour interpretation through guided annotation or feedback workflows.

%% file: sections/8_conclusion.tex
\section{Conclusion}
In this study, we proposed a novel framework for aligning MLLMs with the judgement styles of SLPs. Using joint attention in parent–child interaction as a test case, we explored two key questions: (1) how MLLMs can better \emph{observe} the key interactional cues—namely, \textit{gaze}, \textit{action}, and \textit{vocalisation}; and (2) how MLLMs can be guided to \emph{judge} like SLPs in assessing the quality of these moments.

We first demonstrated that MLLMs, when prompted with structured behavioural segments, can achieve over 90\% similarity with human-generated annotations across action, vocalisation, and gaze fields. This establishes a reliable observation layer that supports higher-level interpretive reasoning. Building upon this, we tested both zero-shot and few-shot approaches for enabling LLMs to assign Strong, Moderate, or Poor labels to segments of joint attention. Our results indicate that few-shot prompting—when seeded with carefully selected, expert-agreed examples—enables LLMs to achieve strong alignment with SLPs who employ more integrative and pattern-based criteria.

These findings show that with targeted prompt engineering and expert-informed training examples, LLMs can begin to approximate expert reasoning processes in real-world social interaction contexts. Our framework provides a step toward more interpretable, human-aligned AI for behavioural evaluation and parent-facing feedback systems.

%% file: main.bib
@article{sunderajan_speech_2019,
	title = {Speech and language delay in children: {Prevalence} and risk factors},
	volume = {8},
	issn = {2249-4863},
	shorttitle = {Speech and language delay in children},
	url = {https://www.ncbi.nlm.nih.gov/pmc/articles/PMC6559061/},
	doi = {10.4103/jfmpc.jfmpc_162_19},
	number = {5},
	urldate = {2025-05-14},
	journal = {Journal of Family Medicine and Primary Care},
	author = {Sunderajan, Trisha and Kanhere, Sujata V.},
	month = may,
	year = {2019},
	pmid = {31198730},
	pmcid = {PMC6559061},
	pages = {1642--1646},
}

@article{robinson2017cdc,
  title={CDC grand rounds: Addressing health disparities in early childhood},
  author={Robinson, Lara R},
  journal={MMWR. Morbidity and mortality weekly report},
  volume={66},
  year={2017}
}

@article{tomasello1986joint,
  title={Joint attention and early language},
  author={Tomasello, Michael and Farrar, Michael Jeffrey},
  journal={Child development},
  pages={1454--1463},
  year={1986},
  publisher={JSTOR}
}

@article{lu2024gpt,
  title={From gpt-4 to gemini and beyond: Assessing the landscape of mllms on generalizability, trustworthiness and causality through four modalities},
  author={Lu, Chaochao and Qian, Chen and Zheng, Guodong and Fan, Hongxing and Gao, Hongzhi and Zhang, Jie and Shao, Jing and Deng, Jingyi and Fu, Jinlan and Huang, Kexin and others},
  journal={arXiv preprint arXiv:2401.15071},
  year={2024}
}

@inproceedings{song2016talklime,
  title={TalkLIME: mobile system intervention to improve parent-child interaction for children with language delay},
  author={Song, Seokwoo and Kim, Seungho and Kim, John and Park, Wonjeong and Yim, Dongsun},
  booktitle={Proceedings of the 2016 ACM International Joint Conference on Pervasive and Ubiquitous Computing},
  pages={304--315},
  year={2016}
}

@inproceedings{hwang2014talkbetter,
  title={TalkBetter: family-driven mobile intervention care for children with language delay},
  author={Hwang, Inseok and Yoo, Chungkuk and Hwang, Chanyou and Yim, Dongsun and Lee, Youngki and Min, Chulhong and Kim, John and Song, Junehwa},
  booktitle={Proceedings of the 17th ACM conference on Computer supported cooperative work \& social computing},
  pages={1283--1296},
  year={2014}
}

@online{dir_floortime,
  title={DIR Floortime},
  author={{The Interdisciplinary Council on Development and Learning}},
  year= {2025},
  url={https://www.icdl.com/dir/floortime},
}

@book{pepper2004talk,
  title={It Takes Two to Talk: A Practical Guide for Parents of Children with Language Delays},
  author={Jan Pepper and Elaine Weitzman},
  year={2004},
  publisher={Hanen Centre},
  note={Shows parents how to help their child communicate and learn language during everyday activities.},
}

@book{sussman1999words,
  title={More Than Words: A Guide to Helping Parents Promote Communication and Social Skills in Children with Autism Spectrum Disorder},
  author={Fern Sussman},
  year={1999},
  publisher={Hanen Centre},
  note={Step by step guide for parents of preschool children with autism spectrum disorder and other social communication difficulties.},
}

@article{o2005barriers,
  title={Barriers to accessing rural paediatric speech pathology services: Health care consumers’ perspectives},
  author={O'Callaghan, Anna M and McAllister, Lindy and Wilson, Linda},
  journal={Australian Journal of Rural Health},
  volume={13},
  number={3},
  pages={162--171},
  year={2005},
  publisher={Wiley Online Library}
}

@inproceedings{chan2017wakey,
  title={WAKEY: assisting parent-child communication for better morning routines},
  author={Chan, Meng-Ying and Lin, Yi-Hsuan and Lin, Long-Fei and Lin, Ting-Wei and Hsu, Wei-Che and Chang, Chia-yu and Liu, Rui and Chang, Ko-Yu and Lin, Min-hua and Hsu, Jane Yung-jen},
  booktitle={Proceedings of the 2017 ACM Conference on Computer Supported Cooperative Work and Social Computing},
  pages={2287--2299},
  year={2017}
}

@inproceedings{kwon2022captivate,
  title={Captivate! contextual language guidance for parent--child interaction},
  author={Kwon, Taeahn and Jeong, Minkyung and Ko, Eon-Suk and Lee, Youngki},
  booktitle={Proceedings of the 2022 CHI Conference on Human Factors in Computing Systems},
  pages={1--17},
  year={2022}
}

@inproceedings{shi2025towards,
  title={Towards Multimodal Large-Language Models for Parent-Child Interaction: A Focus on Joint Attention},
  author={Shi, Weiyan and Le, Hai Viet and Choo, Kenny Tsu Wei},
  booktitle={Proceedings of the Extended Abstracts of the CHI Conference on Human Factors in Computing Systems},
  pages={1--6},
  year={2025}
}

@inproceedings{zheng2024soap,
  title={SOAP. AI: A Collaborative Tool for Documenting Human Behavior in Videos through Multimodal Generative AI},
  author={Zheng, Qingxiao and Rabbani, Parisa and Lin, Yu-Rou and Mansour, Daan and Huang, Yun},
  booktitle={Companion Publication of the 2024 Conference on Computer-Supported Cooperative Work and Social Computing},
  pages={87--90},
  year={2024}
}

@article{mclaughlin2011speech,
  title={Speech and language delay in children},
  author={McLaughlin, Maura R},
  journal={American family physician},
  volume={83},
  number={10},
  pages={1183--1188},
  year={2011}
}

@article{sunderajan2019speech,
  title={Speech and language delay in children: Prevalence and risk factors},
  author={Sunderajan, Trisha and Kanhere, Sujata V},
  journal={Journal of family medicine and primary care},
  volume={8},
  number={5},
  pages={1642--1646},
  year={2019},
  publisher={Medknow}
}

@article{mundy2007individual,
  title={Individual differences and the development of joint attention in infancy},
  author={Mundy, Peter and Block, Jessica and Delgado, Christine and Pomares, Yuly and Van Hecke, Amy Vaughan and Parlade, Meaghan Venezia},
  journal={Child development},
  volume={78},
  number={3},
  pages={938--954},
  year={2007},
  publisher={Wiley Online Library}
}

@inproceedings{lewis2025exploring,
  title={Exploring AI-Based Support in Speech-Language Pathology for Culturally and Linguistically Diverse Children},
  author={Lewis, Aaleyah and Dangol, Aayushi and Suh, Hyewon and Olszewski, Abbie and Fogarty, James and Kientz, Julie A},
  booktitle={CHI Conference on Human Factors in Computing Systems (CHI’25)},
  year={2025},
  organization={ACM New York, NY, USA}
}

@inproceedings{dangol2025want,
  title={“I Want to Think Like an SLP”: A Design Exploration of AI-Supported Home Practice in Speech Therapy},
  author={Dangol, Aayushi and Lewis, Aaleyah and Suh, Hyewon and Hong, Xuesi and Meadan, Hedda and Fogarty, James and Kientz, Julie A},
  booktitle={Proceedings of the 2025 CHI Conference on Human Factors in Computing Systems},
  pages={1--21},
  year={2025}
}

@article{masse2018taking,
  title={Taking PRIDE in your home: Implementing home-based Parent--Child Interaction Therapy (PCIT) with fidelity},
  author={Masse, Joshua J and Quetsch, Lauren Borduin and McNeil, Cheryl B},
  journal={Handbook of parent-child interaction therapy: Innovations and applications for research and practice},
  pages={161--181},
  year={2018},
  publisher={Springer}
}

@article{woodfield2021time,
  title={Time-out with young children: a parent-child interaction therapy (PCIT) practitioner review},
  author={Woodfield, Melanie J and Brodd, Irene and Hetrick, Sarah E},
  journal={International journal of environmental research and public health},
  volume={19},
  number={1},
  pages={145},
  year={2021},
  publisher={MDPI}
}

@inproceedings{choi2025aacess,
  title={AACessTalk: Fostering Communication between Minimally Verbal Autistic Children and Parents with Contextual Guidance and Card Recommendation},
  author={Choi, Dasom and Park, SoHyun and Lee, Kyungah and Hong, Hwajung and Kim, Young-Ho},
  booktitle={Proceedings of the 2025 CHI Conference on Human Factors in Computing Systems},
  pages={1--25},
  year={2025}
}

@article{brock2014statewide,
  title={Statewide assessment of professional development needs related to educating students with autism spectrum disorder},
  author={Brock, Matthew E and Huber, Heartley B and Carter, Erik W and Juarez, A Pablo and Warren, Zachary E},
  journal={Focus on Autism and Other Developmental Disabilities},
  volume={29},
  number={2},
  pages={67--79},
  year={2014},
  publisher={Sage Publications Sage CA: Los Angeles, CA}
}

@article{finke2009all,
  title={“All children can and should have the opportunity to learn”: General education teachers' perspectives on including children with autism spectrum disorder who require AAC},
  author={Finke, Erinn H and Finke, Erinn H and McNaughton, David B and Drager, Kathryn DR},
  journal={Augmentative and Alternative Communication},
  volume={25},
  number={2},
  pages={110--122},
  year={2009},
  publisher={Taylor \& Francis}
}

@article{gadberry2011survey,
  title={A survey of the use of aided augmentative and alternative communication during music therapy sessions with persons with autism spectrum disorders.},
  author={Gadberry, Anita L},
  journal={Journal of music therapy},
  volume={48},
  number={1},
  year={2011}
}

@article{ganz2013impacts,
  title={Impacts of a PECS instructional coaching intervention on practitioners and children with autism},
  author={Ganz, Jennifer B and Goodwyn, Fara D and Boles, Margot M and Hong, Ee Rea and Rispoli, Mandy J and Lund, Emily M and Kite, Elizabeth},
  journal={Augmentative and Alternative Communication},
  volume={29},
  number={3},
  pages={210--221},
  year={2013},
  publisher={Taylor \& Francis}
}

@article{gandhi2023multimodal,
  title={Multimodal sentiment analysis: A systematic review of history, datasets, multimodal fusion methods, applications, challenges and future directions},
  author={Gandhi, Ankita and Adhvaryu, Kinjal and Poria, Soujanya and Cambria, Erik and Hussain, Amir},
  journal={Information Fusion},
  volume={91},
  pages={424--444},
  year={2023},
  publisher={Elsevier}
}

@inproceedings{jain2024vcoder,
  title={Vcoder: Versatile vision encoders for multimodal large language models},
  author={Jain, Jitesh and Yang, Jianwei and Shi, Humphrey},
  booktitle={Proceedings of the IEEE/CVF Conference on Computer Vision and Pattern Recognition},
  pages={27992--28002},
  year={2024}
}

@inproceedings{dos2023composite,
  title={Composite AI for behavior analysis in social interactions},
  author={Dos Santos Melicio, Bruno Carlos and Xiang, Linyun and Dillon, Emily and Soorya, Latha and Chetouani, Mohamed and Sarkany, Andras and Kun, Peter and Fenech, Kristian and Lorincz, Andras},
  booktitle={Companion Publication of the 25th International Conference on Multimodal Interaction},
  pages={389--397},
  year={2023}
}

@inproceedings{whitehead2024generative,
  title={The generative multimodal analysis (gma) methodology for studying socially shared regulation in collaborative learning},
  author={Whitehead, Ridwan and Nguyen, Andy and J{\"a}rvel{\"a}, Sanna},
  booktitle={The International Conference on Learning Analytics \& Knowledge (LAK24)},
  year={2024}
}

@online{straitstimes2024earlyintervention,
  author       = {{The Straits Times}},
  title        = {Long wait times for early intervention push parents of kids with developmental needs to private sector},
  year         = {2024},
  url          = {https://www.straitstimes.com/singapore/long-wait-times-for-early-intervention-push-parents-of-kids-with-developmental-needs-to-private-sector},
  note         = {Accessed: 2025-05-14},
  organization = {The Straits Times}
}

@article{roberts2011parent,
  title={Parent-implemented intervention for children with autism: A systematic review of coaching models},
  author={Roberts, Jacqueline and Kaiser, Ann P},
  journal={Journal of Autism and Developmental Disorders},
  volume={41},
  pages={1027--1041},
  year={2011},
  publisher={Springer}
}

@article{lieneman2017parent,
  title={Parent--child interaction therapy: Current perspectives},
  author={Lieneman, Corey C and Brabson, Laurel A and Highlander, April and Wallace, Nancy M and McNeil, Cheryl B},
  journal={Psychology research and behavior management},
  pages={239--256},
  year={2017},
  publisher={Taylor \& Francis}
}

@article{fu2024video,
  title={Video-mme: The first-ever comprehensive evaluation benchmark of multi-modal llms in video analysis},
  author={Fu, Chaoyou and Dai, Yuhan and Luo, Yongdong and Li, Lei and Ren, Shuhuai and Zhang, Renrui and Wang, Zihan and Zhou, Chenyu and Shen, Yunhang and Zhang, Mengdan and others},
  journal={arXiv preprint arXiv:2405.21075},
  year={2024}
}

@article{yuan2024designing,
  title={Designing Collaborative Technology for Intergenerational Social Play over Distance},
  author={Yuan, Ye and Jin, Qiao and Mills, Chelsea and Yarosh, Svetlana and Neustaedter, Carman},
  journal={Proceedings of the ACM on Human-Computer Interaction},
  volume={8},
  number={CSCW2},
  pages={1--26},
  year={2024},
  publisher={ACM New York, NY, USA}
}

@article{sun2024exploring,
  author = {Sun, Yuling and Chen, Jiaju and Yao, Bingsheng and Liu, Jiali and Wang, Dakuo and Ma, Xiaojuan and Lu, Yuxuan and Xu, Ying and He, Liang},
  title = {Exploring Parent's Needs for Children-Centered AI to Support Preschoolers' Interactive Storytelling and Reading Activities},
  journal={Proceedings of the ACM on Human-Computer Interaction},
  number={CSCW2},
  year = {2024},
}

@article{fiani2024exploring,
  title={Exploring the perspectives of social VR-aware non-parent adults and parents on children's use of social virtual reality},
  author={Fiani, Cristina and Saeghe, Pejman and McGill, Mark and Khamis, Mohamed},
  journal={Proceedings of the ACM on Human-Computer Interaction},
  volume={8},
  number={CSCW1},
  pages={1--25},
  year={2024},
  publisher={ACM New York, NY, USA}
}

@article{nikkhah2024family,
  title={Family Resilience in Care Coordination Technologies: Designing for Families as Adaptive Systems},
  author={Nikkhah, Sarah and Rode, Akash Uday and Kulkarni, Neha Keshav and Mittal, Priyanjali and Mueller, Emily L and Miller, Andrew D},
  journal={Proceedings of the ACM on Human-Computer Interaction},
  volume={8},
  number={CSCW2},
  pages={1--28},
  year={2024},
  publisher={ACM New York, NY, USA}
}

@article{currin2024opportunities,
  title={Opportunities and Challenges in Using Tangible, Teleoperated Voice Agents in Kid-Driven Moments in Play Among Families with Neurodivergent Children},
  author={Currin, Flannery Hope and Kilcoin, Cassidy and Peterman, Kerry and Rector, Kyle and Hourcade, Juan Pablo},
  journal={Proceedings of the ACM on human-computer interaction},
  volume={8},
  number={CSCW1},
  pages={1--25},
  year={2024},
  publisher={ACM New York, NY, USA}
}

@article{su2024hidden,
  title={The Hidden Burden: Encountering and Managing (Unintended) Stigma in Children with Serious Illnesses},
  author={Su, Zhaoyuan and Kamath, Sunil P and Tirakitsoontorn, Pornchai and Chen, Yunan},
  journal={Proceedings of the ACM on Human-Computer Interaction},
  volume={8},
  number={CSCW1},
  pages={1--35},
  year={2024},
  publisher={ACM New York, NY, USA}
}
